\documentclass[aps,reprint,nofootinbib,nobibnotes,notitlepage,superscriptaddress,onecolumn,prd,
 amsmath,amssymb
]{revtex4-2}

\usepackage[caption=false]{subfig}
\usepackage{braket}
\usepackage{float}
\usepackage{lipsum}
\usepackage{graphicx}
\usepackage{dcolumn}
\usepackage{bm}
\usepackage{natbib}
\usepackage{hyperref}
\hypersetup{
	colorlinks = true,
    linkcolor = Red,
    urlcolor  = Red,
    citecolor = Red
}

\usepackage{microtype}

\usepackage{verbatim}
\usepackage[amssymb]{SIunits}
\usepackage{tabularx}
\usepackage[dvipsnames]{xcolor}
\usepackage{wasysym}

\usepackage{tikz}

\def\be{\begin{equation}}
\def\ee{\end{equation}}

\usepackage{color}
\definecolor{darkgreen}{RGB}{0,120,0}

\definecolor{darkgreen}{RGB}{0,120,0}
\newcommand{\resub}[1]{#1}

\newcommand{\delD}[1]{(2\pi)^3\delta_\mathrm{D}\left({#1}\right)}

\newcommand{\av}[1]{\left\langle{#1}\right\rangle} 

\newcommand{\vk}{\vec k}
\newcommand{\hk}{\hat{\vec k}}

\newcommand{\vs}{\vec s}

\newcommand{\Si}{\mathsf{S}^{-1}}
\newcommand{\F}{\mathcal{F}}

\newcommand{\hs}{\hat{\vec s}}
\newcommand{\tjo}[3]{\begin{pmatrix} {#1} & {#2} & {#3}\\ 0 & 0 & 0\end{pmatrix}}
\newcommand{\tj}[6]{\begin{pmatrix} {#1} & {#2} & {#3}\\ {#4} & {#5} & {#6}\end{pmatrix}}

\renewcommand{\L}{\Lambda}
\renewcommand{\P}{\mathcal{P}}

\def\beq{\begin{eqnarray}}
\def\eeq{\end{eqnarray}}

\let\vec\mathbf

\usepackage{empheq}

\def\mnras{\rm{MNRAS}}

\def\aap{\rm{A\&A}}

\begin{document}

\title{Testing Parity Symmetry with the Polarized Cosmic Microwave Background}

\author{Oliver~H.\,E.~Philcox}
\email{ohep2@cantab.ac.uk}
\affiliation{Center for Theoretical Physics, Department of Physics,
Columbia University, New York, NY 10027, USA}
\affiliation{Simons Society of Fellows, Simons Foundation, New York, NY 10010, USA}

\author{Maresuke Shiraishi}
\email{shiraishi\_maresuke@rs.sus.ac.jp}
\affiliation{School of General and Management Studies, Suwa University of Science, Chino, Nagano
391-0292, Japan}

\begin{abstract} 
    \noindent New physics in the early Universe could lead to \textit{parity-violation} in the late Universe, sourcing statistics whose sign changes under point reflection. The best constraints on such phenomena have come from the \textit{Planck} temperature fluctuations; however, this is already cosmic-variance-limited down to relatively small scales, thus only small improvements are expected in the future. Here, we search for signatures of parity-violation in the polarized CMB, using the \textit{Planck} PR4 $T$- and $E$-mode data. We perform both a simulation-based blind test for any parity-violating signal at $\ell<518$, and a targeted search for primordial $U(1)$ gauge fields (and the amplitudes of a generic collapsed model) at $\ell<2000$. In all cases, we find no evidence for new physics, with the model-independent test finding consistency with the FFP10/\textsc{npipe} simulation suite at $(-)0.4\sigma$, and the gauge field test constraining the fractional amplitude of gauge fields during inflation to be below $6\times 10^{-19}$ at $95\%$ confidence level for a fiducial model. The addition of polarization data can significantly improve the constraints, depending on the particular model of primordial physics, and the bounds will tighten significantly with the inclusion of smaller-scale information.
\end{abstract}

\maketitle

\section{Introduction \& Motivation}\label{sec: intro}
\noindent By measuring the large-scale temperature anisotropies present in the cosmic microwave background (CMB), satellites such as \textit{Planck} and WMAP have transformed our view of the primordial Universe \citep[e.g.,][]{WMAP:2003elm,WMAP:2010qai,2020A&A...641A...6P,Planck:2019kim,Planck:2018jri}. To some extent, this mission is already complete: due to the cosmic variance limit, future will not yield additional information on the Universe from large scales ($\ell\lesssim 2000$). Polarization, on the other hand, provides new clues to terrestrial cosmic detectives. In \textit{Planck}, the cosmic variance limit impacts $E$-modes only on the largest scales, and investigations into $B$-modes are hampered only by noise and lensing effects. As such, precision analyses of polarization will take the forefront in the search for fundamental physics from the CMB; indeed, this is already occurring with observatories such as ACT, SPT, and BICEP/Keck \citep{ACT:2020gnv,SPT-3G:2022hvq,BICEPKeck:2022mhb}.

To date, a wealth of new physics has been searched for in CMB temperature and polarization fluctuations, probing both the scalar and tensor sectors in the early and late Universe (see \citep{Planck:2018jri,Planck:2019kim} and citations therein). A particularly novel example is that of parity-violation; physics that changes sign under point-reflection. Although there exists a range of physical mechanisms to generate such effects \citep{Wu:1957my,CyrilCS,PhysRev.104.254,Shiraishi:2016mok,Alexander:2006mt,Gluscevic:2010vv,Shiraishi:2013kxa,Bartolo:2014hwa,Shiraishi:2012sn,Shiraishi:2011st,Lue:1998mq,Cabass:2022rhr,Cabass:2022oap,Alexander:2009tp,Alexander:2016hxk,Sakharov:1967dj,Liu:2019fag,Abedi:2018top}, their signatures are difficult to probe in nature; if sourced by scalars, they appear only in four-point correlation functions and beyond (\citep{Shiraishi:2016mok}, see also \citep{Lue:1998mq,Gluscevic:2010vv,Cahn:2021ltp,Coulton:2023oug}).\footnote{Physics sourced by gravitational waves, on the other hand, can be observed in lower-order statistics of $T$- and $E$-modes, e.g. bispectra with odd $\ell_1+\ell_2+\ell_3$ \citep{Shiraishi:2014ila,Planck:2015zfm,Planck:2019kim,Philcox:2023xxk}.} Difficult does not imply impossible however, and recent works have reported tentative signatures of parity-violation in the late-time distribution of galaxies (\citep{Philcox:2022hkh,Hou:2022wfj}, using the approach described in \citep{Cahn:2021ltp}) and in studies of CMB birefringence \citep{Eskilt:2022cff,Diego-Palazuelos:2022dsq,Eskilt:2022wav}, though the latter requires late-time vector sourcing, and will not be considered in this work. 

If the purported signal in the galaxy trispectrum were physical and primordial, what would it imply for CMB observations?\footnote{See \citep{Cabass:2022oap} for a discussion of potential late-time sources and their (in)viability.} As discussed in \citep{PhilcoxCMB,Shiraishi:2016mok}, parity-violation in the scalar sector imprints itself in projected two-sphere observables via a non-vanishing $T$- or $E$-mode trispectrum with odd $\ell_1+\ell_2+\ell_3+\ell_4$. Geometrically, this can be thought of as a difference between trispectra of points with left- and right-handed chiralities. In the flat-sky and instantaneous-recombination limit, such effects vanish since observations are confined to a plane, and reflections become equivalent to rotations; in practice, this effect does not limit us, since it is relevant only when \textit{all} sides of the tetrahedron (including its diagonals) are small (\textit{i.e.}\ at high $\ell$); a regime usually orthogonal to the signals of interest. 

In \citep{PhilcoxCMB}, the trispectrum of the CMB temperature fluctuations was used to search for scalar parity-violation. Despite the analysis containing $\approx 250\times$ more modes than the large-scale structure equivalent, no signal was found in a blind analysis, casting doubt on any primordial explanation of the former results. Here, we extend the analysis to the polarization sector (and additionally use updated \textit{Planck} PR4 data \citep{Planck:2020olo,Tristram:2020wbi,Rosenberg:2022sdy}), making use of the new trispectrum estimators described in \citep{Philcox:2023psd} (building on \citep{Philcox:2023uwe}). Noting that $E$-modes also trace primordial scalar physics, this can lead to a significant boost in signal-to-noise (depending on the model in question), particularly due to the large number of $T$- and $E$-mode trispectra that can be wrought ($2^4$ combinations) (cf.\,\citep{Planck:2019kim} for polarized bispectra). Our purpose is not to cast further aspersions on the large-scale structure model; rather, we wish to place novel bounds on little-known physics with high-precision data, which can be greatly improved with upcoming higher-resolution polarization experiments. We will further show the utility of performing targeted searches for models of interest, by constraining the gauge-field model described in \citep{Shiraishi:2016mok}. This peaks in collapsed configurations, and its amplitude was seen at $\approx 2\sigma$ hint in the temperature data \citep{PhilcoxCMB}.

\vskip 4pt 

The remainder of this paper is structured as follows. In \S\ref{sec: data} we discuss the \textit{Planck} data-set, before describing the details of our trispectrum estimation pipeline in \S\ref{sec: estimation}. \S\ref{sec: results-blind}\,\&\,\S\ref{sec: results-gauge} present the main results of our analysis: general and targeted searches for primordial non-Gaussianity, before we conclude by discussing their implications are discussed in \S\ref{sec: summary}. Appendix \ref{app: model} outlines the derivativation of the gauge-field model used in \S\ref{sec: results-gauge}, whilst Appendix \ref{app: consistency} presents various consistency checks of our estimators. 

\vskip 4pt
\paragraph*{Conventions} Throughout the work, we define the (connected) trispectrum measured from a field $a_{\ell m}^X$ as:
\beq\label{eq: Tl-def}
    \av{\prod_{i=1}^4a_{\ell_im_i}^{X_i}}_c = \sum_{LM}(-1)^Mw^{L(-M)}_{\ell_1\ell_2m_1m_2}w^{LM}_{\ell_3\ell_4m_3m_4}t^{X_1X_2,X_3X_4}_{\ell_1\ell_2,\ell_3\ell_4}(L)+\text{23 perms.}
\eeq
where $X_i\in\{T,E,B\}$, and $L,M$ parametrize the internal leg of the tetrahedron \citep{Regan:2010cn,Philcox:2023psd}. This defines the weighting functions
\beq
    w^{LM}_{\ell_1\ell_2m_1m_2} = \sqrt{\frac{(2\ell_1+1)(2\ell_2+1)(2L+1)}{4\pi}}\tj{\ell_1}{\ell_2}{L}{-1}{-1}{2}\tj{\ell_1}{\ell_2}{L}{m_1}{m_2}{M},
\eeq
which are akin to Gaunt integrals, but with modified spins. Trispectra containing an even number of $B$-modes are said to be \textit{parity-even} if $\ell_1+\ell_2+\ell_3+\ell_4$ is even, and \textit{parity-odd} if $\ell_1+\ell_2+\ell_3+\ell_4$ is odd; the situation is reversed if there are an odd number of $B$-modes. We assume a fiducial cosmology defined by the \textit{Planck} 2018 parameters: $\omega_b = 0.022383$, $\omega_{cdm} = 0.12011$, $h = 0.6732$,  $\tau = 0.0543$, $\log 10^{10}A_s = 3.0448$, $n_s = 0.96605$, for a single massive neutrino species with $m_\nu = 0.06\,\mathrm{eV}$ \citep{2020A&A...641A...6P}. To avoid confirmation bias, the pipeline discussed below was developed using only mock data, with the \textit{Planck} results added as a final step.

\section{Data}\label{sec: data}
\noindent Throughout this work, we utilize the fourth data release (PR4) from the \textit{Planck} satellite \citep{Planck:2020olo,Rosenberg:2022sdy,Tristram:2020wbi}, an upgrade from the \textit{Planck} 2018 (PR3) data used in the previous work \citep{PhilcoxCMB}.\footnote{These have been made publicly available on NERSC, as described at \href{https://portal.nersc.gov/project/cmb/planck2020/}{portal.nersc.gov/project/cmb/planck2020/}.} This features improved treatment of polarization, noise, and large-scale systematics, which are particularly relevant to this study. In addition to the data, we use 600 end-to-end FFP10 simulations \citep{Planck:2020olo}, with both data and mocks processed with the \textsc{npipe} pipeline, using the \textsc{sevem} component separation algorithm.\footnote{This is preferred over the alternative \textsc{commander} algorithm, since the latter does not contain joint temperature and polarization simulations in PR4.} These maps are smoothed to both $N_{\rm side}=256$ and $N_{\rm side}=1024$ resolution in \textsc{HealPix} \citep{Gorski:2004by}, facilitating different types of study.

\begin{figure}
    \centering
    \includegraphics[width=0.48\linewidth]{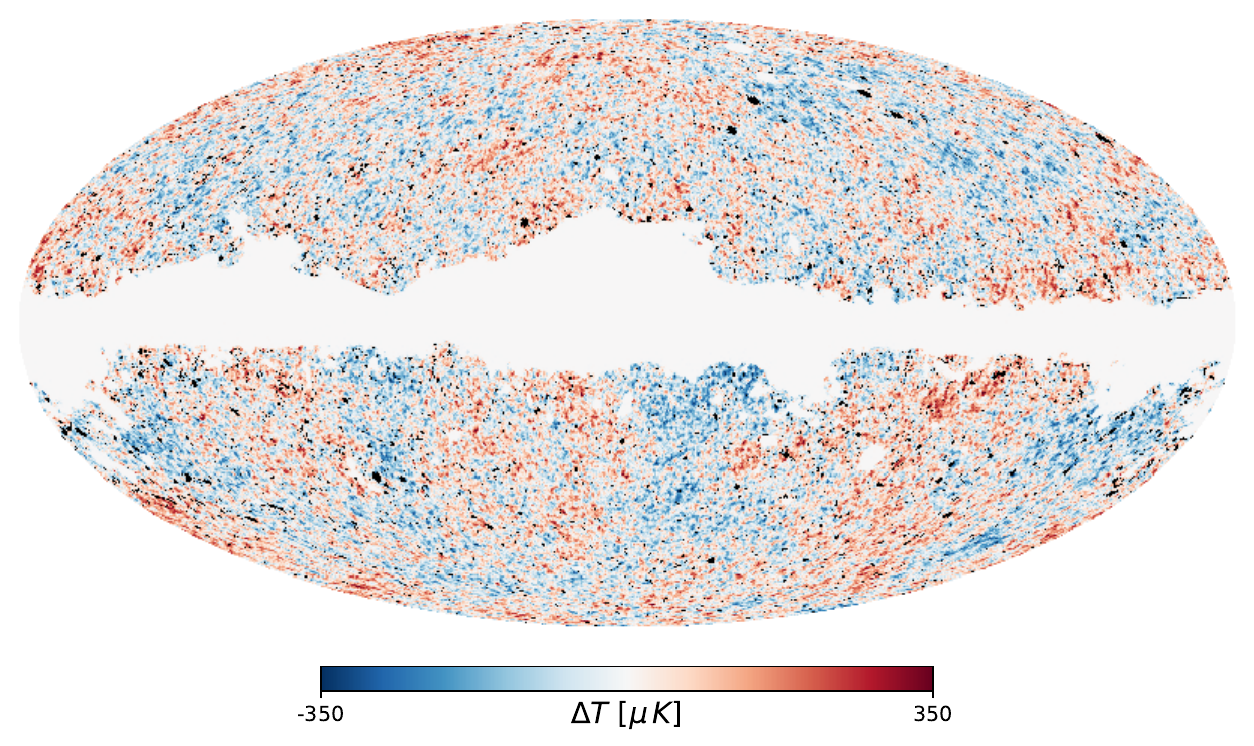}
    \includegraphics[width=0.48\linewidth]{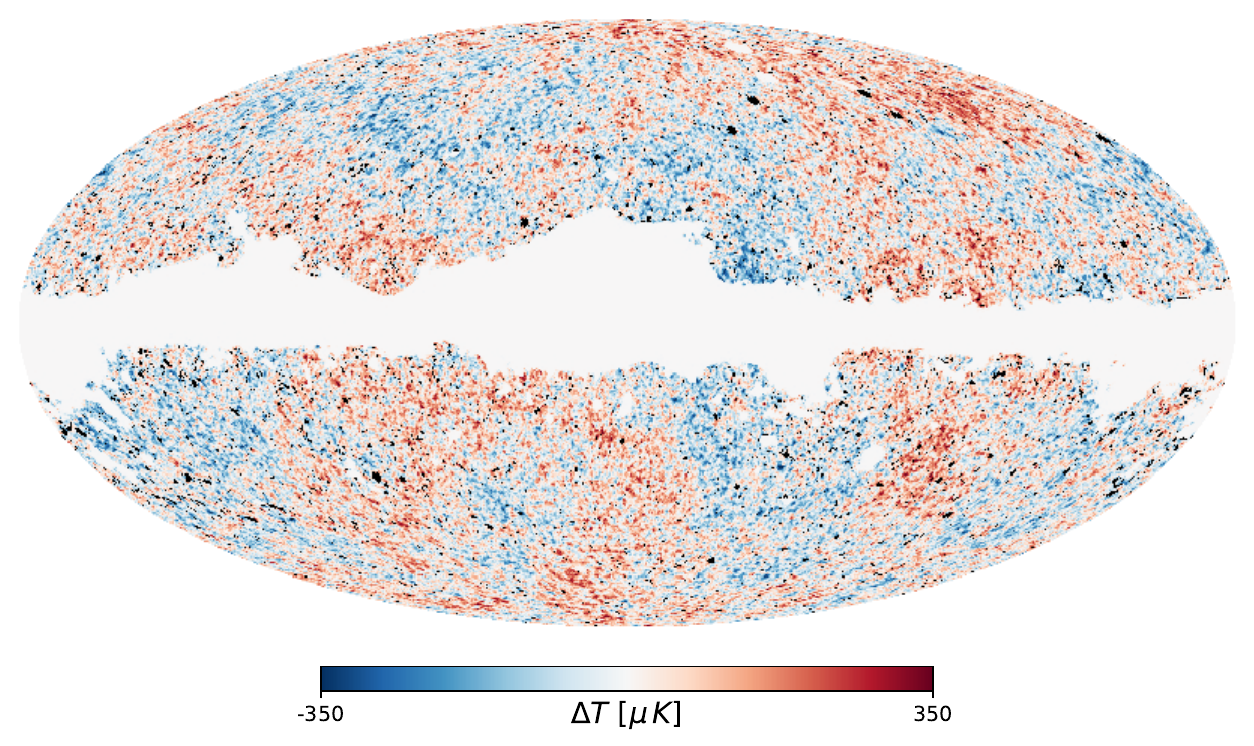}
    \includegraphics[width=0.48\linewidth]{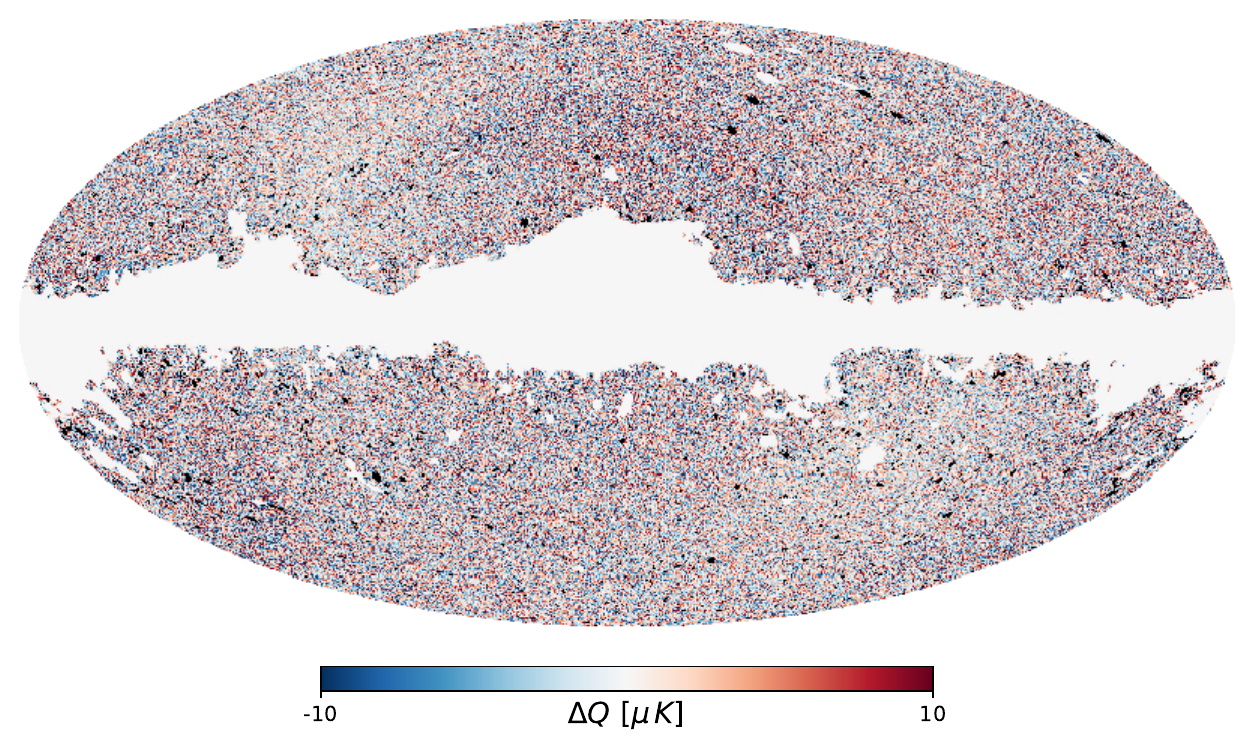}
    \includegraphics[width=0.48\linewidth]{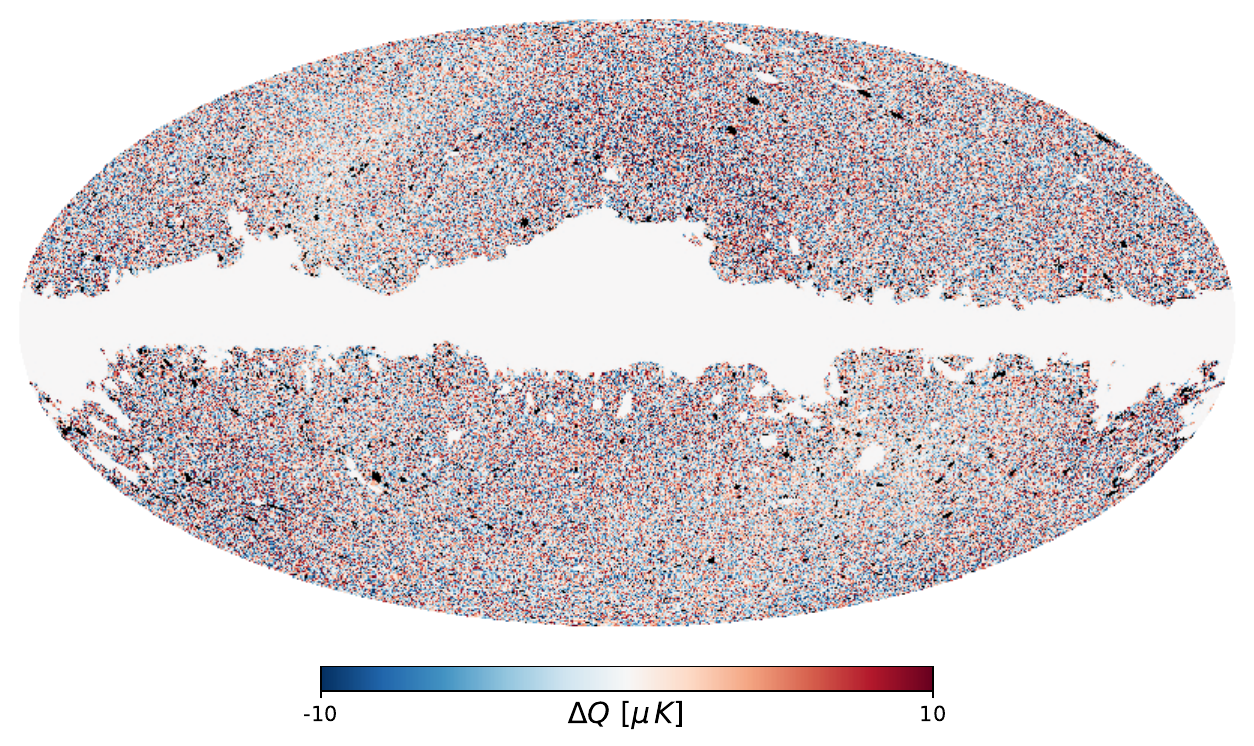}
    \includegraphics[width=0.48\linewidth]{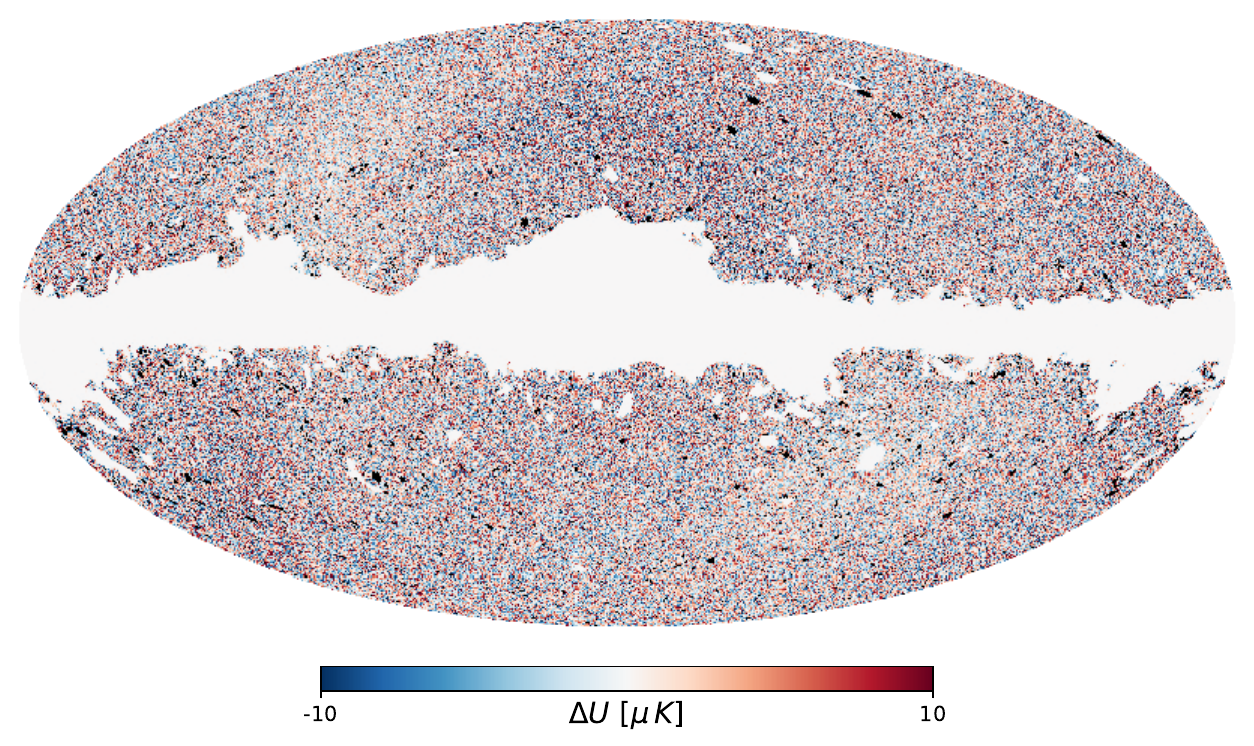}
    \includegraphics[width=0.48\linewidth]{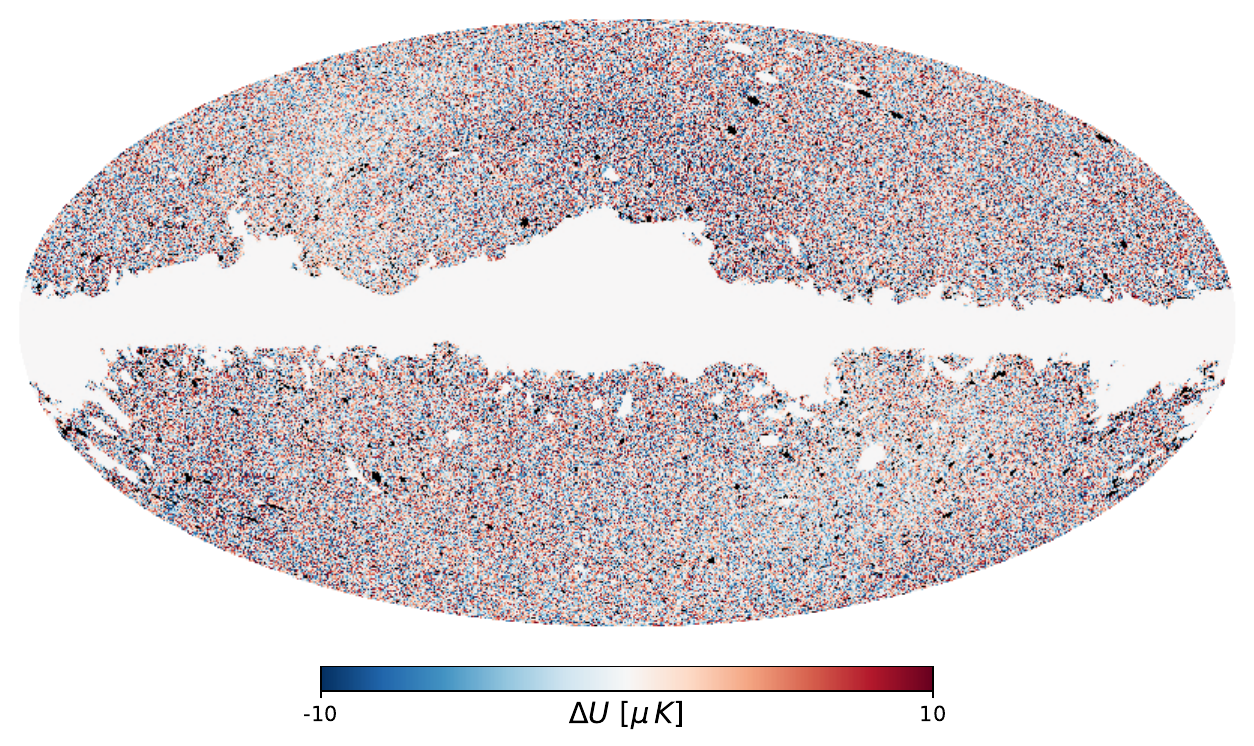}
    \caption{Comparison between the \textit{Planck} data (left) and a single FFP10 simulation (right). The three plots show $T$, $Q$, and $U$ maps (in $\mu \rm{K}$ units), all of which are used in the analysis. Black regions show pixels that are inpainted in our analysis, whilst those in white are masked. We note that the FFP10 simulations capture the large-scale (systematic) non-Gaussian features in the polarized maps.}
    \label{fig: maps}
\end{figure}

When computing trispectra, we filter the data via a linear operator $\Si$. This firstly inpaints small holes and point sources in the map (containing less than $10(N_{\rm side}/128)^2$ empty pixels) via a linear scheme \citep[cf.][]{Gruetjen:2015sta}, before applying the \textit{Planck} 2018 common component-separation mask \citep{Planck:2018yye}, smoothed to $10'$ resolution, which retains $\approx 78\%$ of the sky.\footnote{Note that the mask is here included in the $\Si$ weights, rather than the estimator itself \citep[cf.,][]{Philcox:2023psd,PhilcoxCMB}. This leads to reduced leakage between bins and does not induce bias. Our approach matches \citep{2015arXiv150200635S} for the scalar case.} Secondly, we filter the data in harmonic space via a diagonal-in-$\ell$ weighting $S^{XY}_\ell$, which is the sum of a theoretical (parity-even) power spectrum and a noise spectrum, extracted from the PR4 half-mission maps at $N_{\rm side}=2048$. We assume parity-even noise ($N^{EB} = 0$, $N^{EE}=N^{BB}$), and ignore any temperature-to-polarization leakage ($N^{TE}=N^{TB}=0$). Finally, we include the beam (both in the filtering and the estimator); this matches the PR3 release, with the addition of an \textsc{npipe} polarization transfer function, required to remove spurious effects in the $\ell<40$ $E$-mode (and $B$-mode) power spectrum \citep{Planck:2020olo}. The latter is computed from the cross-spectra of 600 FFP10 simulations (filtered as above) with the input realizations. Whilst our filtering is not strictly optimal (since we do not fully account for off-diagonal harmonic-space correlations induced by the mask, nor translation-invariant noise), such effects are expected to be small except on the largest scales (and do not bias the results) \citep{Philcox:2023psd}. In \S\ref{sec: results-blind}, we will additionally make use of Gaussian random field simulations; these are generated as masked Gaussian realizations of the $S_{\ell}^{XY}$ power spectra.

In Fig.\,\ref{fig: maps}, we plot the masked \textit{Planck} temperature and polarization maps, alongside the first FFP10 simulation. The agreement between simulation and data is excellent (though the underlying CMB realization differs); importantly, in both cases, we see non-Gaussian signatures in the polarized maps, exhibiting a quadrupole-like variation in power. This can be attributed to residual foreground effects \citep{Planck:2020olo}, \resub{and could potentially give a spurious signal in the polarized trispectrum. Since this is present in both simulations and data, we can verify our trispectrum pipeline on the FFP10 simulations to check for such effects, and, if necessary (which is not true in this work), subtract their mean value.}

\section{Trispectrum Estimation}\label{sec: estimation}
\noindent Given the \textit{Planck} and FFP10 simulations, the experimental beam, and the $\Si$ weighting, we compute the observed trispectra using the \href{https://github.com/OliverPhilcox/PolyBin}{\textsc{PolyBin}} code \citep{PolyBin}, described in \citep{Philcox:2023uwe,Philcox:2023psd}, loosely based on the template approaches of \citep{2011MNRAS.417....2S,2015arXiv150200635S}.\footnote{Publicly available at \href{https://github.com/oliverphilcox/PolyBin}{github.com/OliverPhilcox/PolyBin}.} This is a quasi-optimal estimator of the full-sky trispectrum, \resub{which takes the form}
\beq
    t^u(\vec b,\beta) = \sum_{\vec b',\beta',u'}\F^{-1,uu'}(\vec b,\beta;\vec b',\beta')t^{{\rm num},u'}(\vec b',\beta'),
\eeq
comprising two contributions: a data-dependent numerator, $t^{\rm num}$, and a data-independent normalization, $\mathcal{F}$ \resub{(both of which can be efficiently computed). Here $t^u(\vec b,\beta)$ is the trispectrum in $\ell$-bins $\vec b\equiv \{b_1,b_2,b_3,b_4\}$, \resub{internal $L$-bin $\beta$} and fields $u\equiv\{X_1,X_2,X_3,X_4\}$ (for $X_i\in\{T,E\}$). The numerator is the sum of three contributions:}
\beq\label{eq: t-num}
    t^{{\rm num},u}(\vec b,\beta) &=& \frac{1}{24}\frac{\partial\av{d^id^jd^kd^l}}{\partial t^u(\vec b,\beta)}\bigg\{[\Si d]^i[\Si d]^j[\Si d]^k[\Si d]^l\\\nonumber
    &&\qquad\qquad\qquad\qquad\,-\,6\av{[\Si d]^i[\Si d]^j}[\Si d]^k[\Si d]^l\\\nonumber
    &&\qquad\qquad\qquad\qquad\,+\,3\av{[\Si d]^i[\Si d]^j}\av{[\Si d]^k[\Si d]^l}\bigg\},
\eeq
\resub{where $i,j,k,l$ run over all pixels and fields in the dataset, $d$ (assuming Einstein summation). Only the first term is strictly required for the parity-odd estimators (which involves four copies of the data, or $d_i\equiv a^u_{\ell m}$), the other terms remove Gaussian contributions to the parity-even estimator and can reduce variance. The normalization matrix, $\F$, is set such that} the estimator is unbiased for any weighting scheme $\Si$, and \resub{can ameliorate mask-induced leakage between bins, polarizations, and parities, as well as correlations between $T$ and $E$, inpainting and beam effects. Explicitly, it is given by}
\beq\label{eq: t-fish}
    \F^{uu'}(\vec b,\beta;\vec b',\beta') &=& \frac{1}{24}\frac{\partial\langle{d^id^jd^kd^l\rangle}}{\partial t^u(\vec b,\beta)}\Si_{ii'}\Si_{jj'}\Si_{kk'}\Si_{ll'}\frac{\partial\langle{d^{i'}d^{j'}d^{k'}d^{l'}\rangle}}{\partial t^{u'}(\vec b',\beta')}.
\eeq
In the limit of an ideal weighting scheme and a Gaussian dataset (\textit{i.e.}\ $\Si$ equal to the inverse of the pixel covariance, \resub{$\av{dd^{\rm T}}$}), the estimators are optimal, such that their covariance is equal to $\F^{-1}$ \citep{Philcox:2023psd,Hamilton:1999uw}. \resub{Although verification of the estimators on parity-violating simulations is beyond the scope of this work, the trispectrum estimators have been extensively tested in \citep{Philcox:2023uwe,Philcox:2023psd}, most notably by assessing their idealized variances and application to Gaussian simulations.}

Here, we perform two tests for parity-violation. Firstly, we compute the \resub{\textit{Planck}} trispectra across a broad range of bins and configurations, working at modest $\ell_{\rm max}$ and $N_{\rm side}=256$ (echoing \citep{PhilcoxCMB}). We compute the trispectra for each non-trivial combination of $T$ and $E$-modes ($TTTT$, $TTTE$, $TTEE$, $TETE$, $TEEE$, $EEEE$, with appropriate restrictions on bins to avoid degeneracies), noting that $B$-modes are not sourced by scalar physics at leading order, and any mask-induced leakage is small (though we consider this explicitly in Appendix \ref{app: consistency}). For this purpose, we use $N_\ell=7$ linear bins equally-spaced in $\ell^{2/5}$ with $\ell\in[2,518]$ \resub{(somewhat coarser than \citep{PhilcoxCMB} to temper the increase in dimensionality caused by the addition of polarization), such that the signal-to-noise is roughly constant in each bin (according to the primordial scalings of \citep{Kalaja:2020mkq})}.\footnote{The bin-edges are given by $\{2,   4,  22,  62, 128, 223, 352, 518\}$.} The full trispectrum is specified by four $\ell$-bins (for the four sides) and one diagonal $L$-bin (cf.\,\ref{eq: Tl-def}), giving a total of $N_{\rm bin}=3130$ bins in the datavector (across all polarizations), or $2386$ when the first $\ell$-bins are dropped due to reduce leakage from unmeasured bins. By default, we do not include any parity-even modes in the analysis; this assumption is verified in Appendix \ref{app: consistency}. \resub{Whilst finer $\ell$-bins and larger $\ell_{\rm max}$ would facilitate (somewhat) higher precision analyses, the corresponding trispectra become significantly more computationally expensive. The above values were chosen to balance these factors.}

Secondly, we perform a targeted analysis of a specific gauge-field model (described in \citep{Shiraishi:2016mok} and \S\ref{sec: results-gauge}), whose signatures primarily lie in collapsed trispectra with $L\lesssim 5$) In this case, we may extend to higher resolution ($N_{\rm side}=1024$), since we require far fewer trispectrum bins to characterize the signal (with the number of collapsed bins scaling as $\ell_{\rm max}^2$, rather than $\ell_{\rm max}^5$ for general bins). In this case, we use $N_\ell=7$ linear bins logarithmically-spaced in $[2,2000]$ \resub{(chosen to allow easy exploration of the $\ell_{\rm max}$-dependence with modest computation resources)}, but restrict to $L\leq 10$ for the internal leg.\footnote{The bin-edges are given by $\{2,    5,   10,   30,  100,  300, 1000, 2000\}$.} This yields $N_{\rm bin}=484$ bins in total, or 342 after subtraction of the smallest $\ell$-bin.

\resub{In practice, the trispectrum numerator given in \eqref{eq: t-num}} is computed from the data set $d$ via a series of \resub{spin-weighted} spherical harmonic transforms, as described in \citep{Philcox:2023psd}, \resub{with the expectation terms computed via Monte Carlo summation using the FFP10 simulations}.\footnote{Note that the error induced by an incorrectly assumed fiducial cosmology starts at second order in $C_\ell^{\rm fid}-C_\ell^{\rm true}$ \resub{and is heavily suppressed by the restriction to parity-odd modes. The fiducial cosmology also appears in the weighting scheme; this cancels in the narrow-bin limit and cannot induce bias.}} Here, we use $100$ simulations for the blind tests, or $10$ for the targeted analysis, noting that such terms are sourced only by mask-induced leakage (though can inflate the error-bars on large scales). Computation of the disconnected terms is usually rate-limiting, and scales as $N_{\rm sim}N_{\ell}^2$ for arbitrary trispectra of $N_{\rm sim}N_{\ell}$ for collapsed trispectra. Secondly, the normalization matrix, $\mathcal{F}$, is computed also via Monte Carlo methods \resub{(in particular, the Girard-Hutchinson stochastic trace estimator \citep{girard89,hutchinson90})}, making use of $N_{\rm mc}=10$ Gaussian random field simulations generated at the fiducial cosmology. This scales linearly with $N_{\rm bin}$ and $N_{\rm mc}$, though memory is usually the limiting factor. \resub{As shown in \citep{Philcox:2023psd}, $N_{\rm mc}=10$ realizations facilitates good convergence; for the analysis below, we validate this by computing also trispectra with $N_{\rm mc}=5$, which differ by only $\mathcal{O}(0.01\sigma)$ inducing a bias in the parity-violating $\chi^2$ statistic (see \S\ref{sec: results-blind}) below $2\%$.}

In the above configurations, each blind-test numerator required $\approx 90$ CPU-hours to compute (involving $\approx 60\,000$ harmonic transforms), whilst the normalization matrix required $\approx 450$ CPU-hours (and $\approx 80\,000$ harmonic transforms) per realization and $1\,\mathrm{TB}$ memory. For the collapsed trispectra, each numerator required $\approx 30$ CPU-hours ($6\,000$ transforms), whilst each normalization realization required $\approx 2000$ CPU-hours ($15\,000$ transforms). We note that (a) the numerator run-time could be significantly reduced by using fewer disconnected simulations (whose impact should be small), and (b) the Fisher computation is not optimally parallelized, since memory considerations usually imply exclusive node usage.

\section{Results: Blind Test}\label{sec: results-blind}
\noindent We begin by performing a \resub{``blind test''} for parity-violation, \resub{whereupon we search for a (model-agnositc) excess in the binned trispectrum relative to the expected noise fluctuations (following the methodology of \citep{PhilcoxCMB})}. To this end, we compute parity-odd trispectra with $\ell\in[2,518]$ for the \textit{Planck} data, in addition to 600 FFP10 simulations and 1000 Gaussian random field realizations. 

\subsection{Simulations \& Covariances}
\begin{figure}
    \centering
    \includegraphics[width=0.49\textwidth]{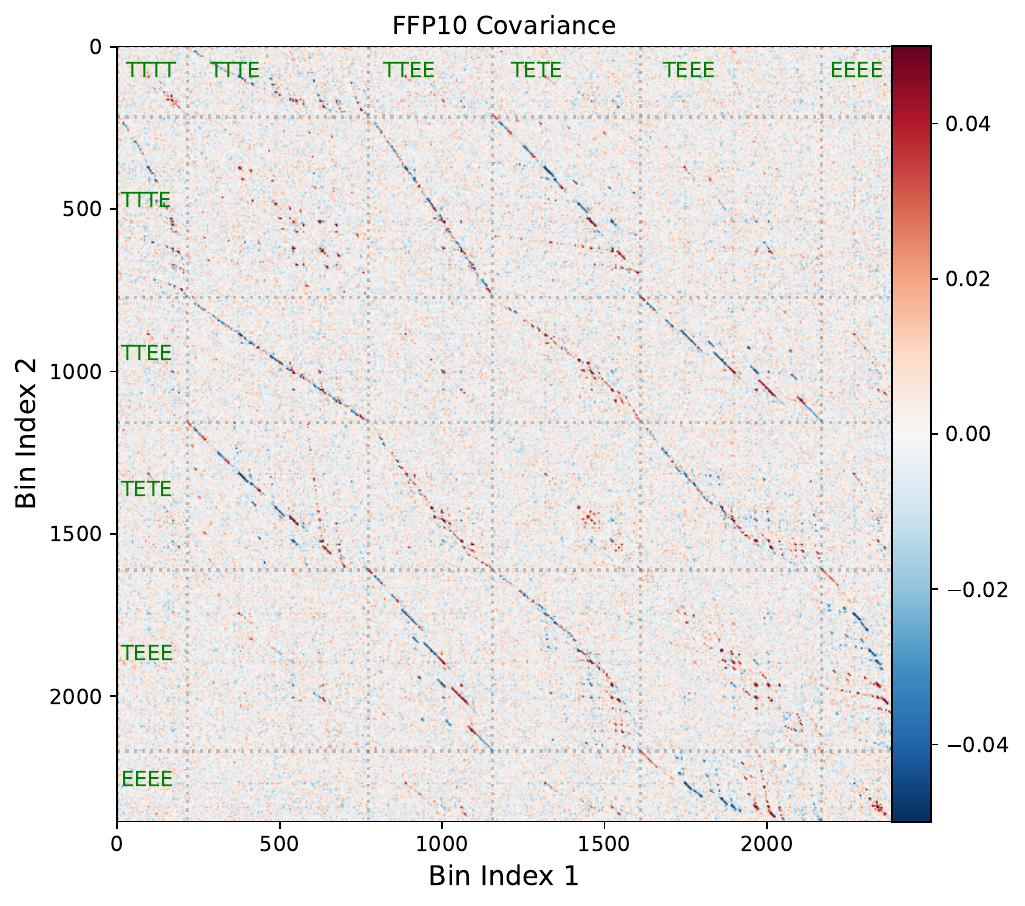}
    \includegraphics[width=0.49\textwidth]{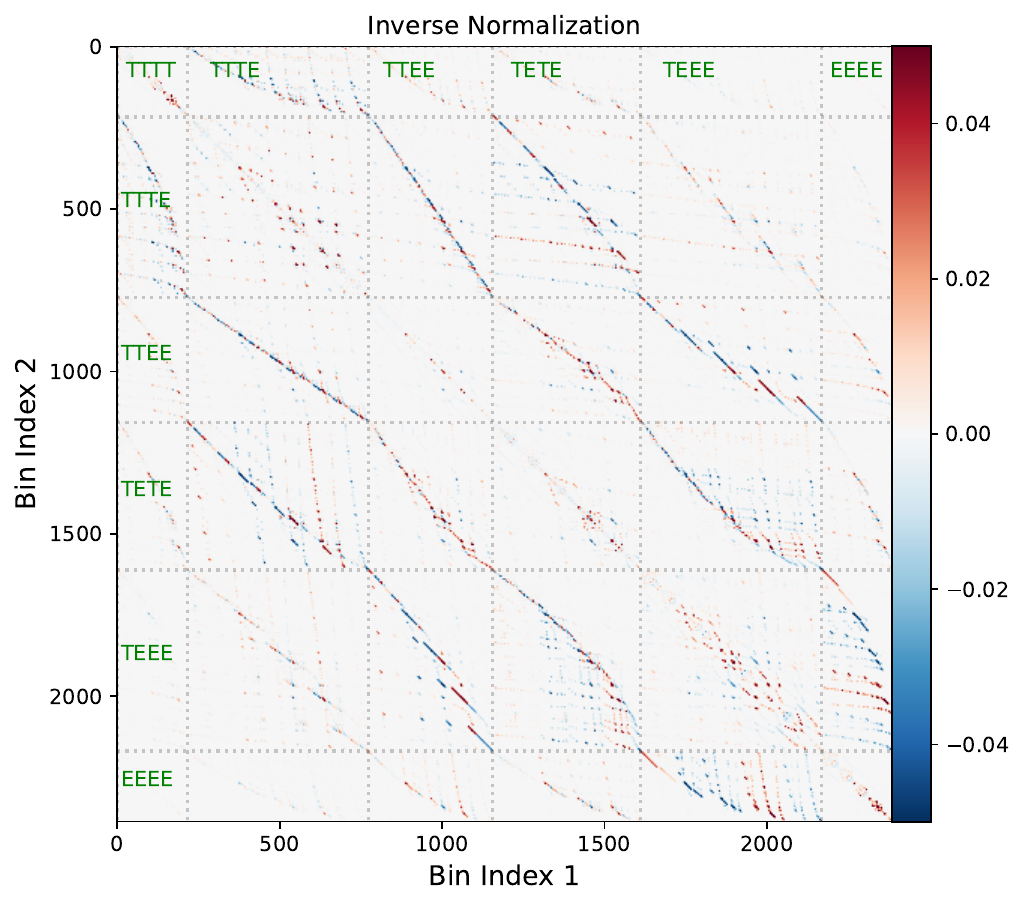}
    \caption{\resub{Comparison of empirical (left) and theoretical (right) correlation matrices for the trispectrum, $t^u(\vec b,\beta)$. The empirical matrices are computed using 600 FFP10 simulations, whilst the theoretical prediction is given by the inverse trispectrum normalization matrix, itself obtained via Monte Carlo methods, using only the survey mask and weighting scheme.} In each panel, we stack all $2386$ $T$- and $E$-mode trispectrum measurements with $4\leq\ell<518$ into one-dimension (ordered from low- to high-$\ell$), \resub{separating out the six non-trivial combinations of fields, which are marked in green. For example, the submatrix in the second column and third row contains the $TTTE\times TTEE$ covariance.} Note that different correlators contain different numbers of bins, due to labelling degeneracies (\textit{i.e.}\ restricting to $\ell_1<\ell_2$ for fields $X_1\neq X_2$). We subtract the (unit) diagonal of the matrix for clarity, and truncate the scale to $\pm5\%$. The empirical correlation matches the inverse normalization to high precision (up to noise), implying that the latter can be used to decorrelate the measured data, \resub{as shown in Fig.\,\ref{fig: tau-corr}}.}\label{fig: corr}
\end{figure}

\noindent A necessary ingredient in the estimators is the normalization matrix, $\mathcal{F}$; as discussed above, this is equal to the inverse covariance matrix of the trispectrum if the estimator is optimal. In Fig.\,\ref{fig: corr}, we test this by comparing the structure of $\mathcal{F}^{-1}$ to the empirical correlation matrix derived from the FFP10 trispectra. Both matrices have a complex structure, which is off-diagonal at the $\mathcal{O}(10\%)$ level. This arises due to three main effects: (1) intrinsic correlations between $T$- and $E$-modes (which can be large); (2) mask-induced leakage between neighboring bins (which are not necessarily neighboring in our flattened trispectrum representation); (3) intrinsic correlations due to a degeneracy in the trispectrum diagonal definition. Clearly, these effects must be included in any analysis pipeline; here, we find that they are very well described by the inverse normalization matrix, $\mathcal{F}^{-1}$. 

\begin{figure}
\centering
\begin{minipage}{.43\textwidth}
  \includegraphics[width=0.99\textwidth]{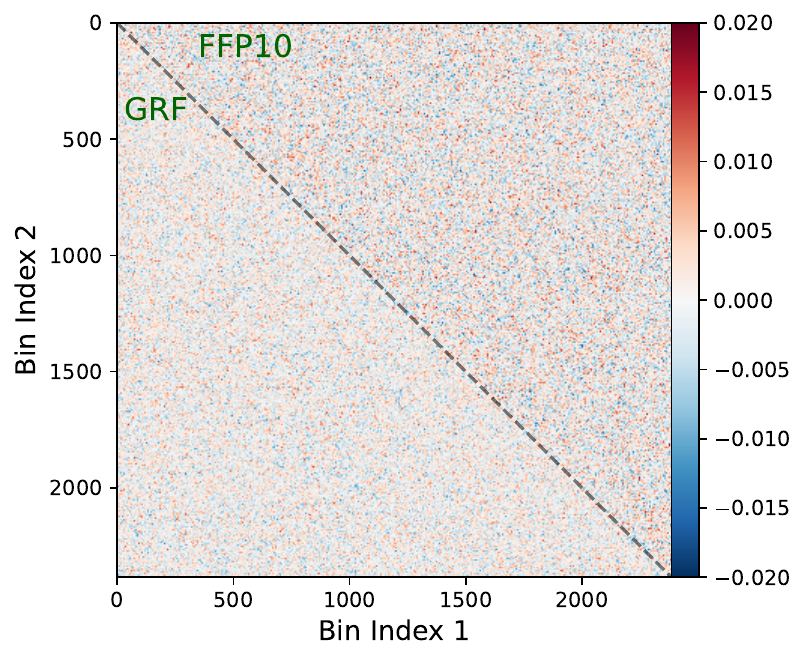}
    \caption{Correlation matrix of the rescaled trispectrum statistic $\tau$, \resub{defined by \eqref{eq: tau-def}}. This is obtained using both 600 FFP10 simulations (upper left) and 1000 GRF simulations (bottom right). \resub{Unlike the unscaled trispectrum correlations shown in Fig.\,\ref{fig: corr}, we find no evidence for off-diagonal contamination in the $\tau$ covariances, with both matrices} consistent with noise. As in Fig.\,\ref{fig: corr}, we stack all bins into one dimension and subtract the leading diagonal (which is unity only if the estimator is optimal).}\label{fig: tau-corr}
\end{minipage}%
\hspace{10pt}
\begin{minipage}{.54\textwidth}
  \centering
  \includegraphics[width=0.9\textwidth]{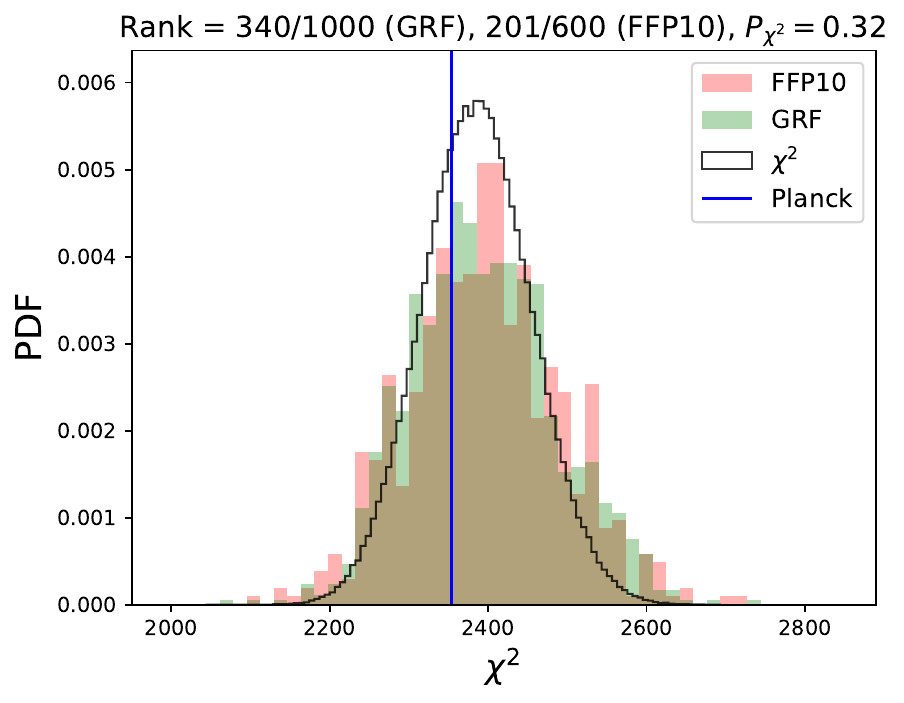}
    \caption{Probability of parity-violation in the large-scale ($4\leq \ell <518$) \textit{Planck} PR4 dataset. We compare the $\chi^2$ statistic 
    of \eqref{eq: chi2-def} measured in \textit{Planck} to empirical distributions from FFP10 and GRF simulations (red and green respectively). We find excellent agreement between \textit{Planck} results and the simulation histograms, with frequentist ranks and $\chi^2$ probabilities-to-exceed shown in the title. Both empirical distributions exhibit slightly heavier tails than the theoretical distribution, implying that classical $\chi^2$-analyses are insufficient for modeling the trispectrum noise properties. The frequentist detection probability for \textit{Planck} corresponds to $-0.4\sigma$.}
    \label{fig: chi2-results}
\end{minipage}
\end{figure}

The above discussion motivates us to define the rescaled trispectra:
\beq\label{eq: tau-def}
    \resub{\tau^u(\vec b,\beta) = \left[\left(\mathcal{F}^{1/2}\right)^{\rm T}t\right]^u(\vec b,\beta)},
\eeq
\resub{where $\mathcal{F}^{1/2}$ is the Cholesky factorization of the (binned) normalization matrix $\mathcal{F}$ (defined in \ref{eq: t-fish})}.\footnote{Note that the lowest $\ell$-bins are removed from the datavector \textit{before} $\tau$ is computed to ensure unbiasedness.} Noting that structure of $\mathcal{F}^{-1/2}$ is approximately equal to that of the covariance of $t$ (up to one-dimensional rescalings), this decorrelates the various trispectrum components, which significantly simplifies analysis \citep{Hamilton:1999uw,PhilcoxCMB}. As a further verification, we plot the correlation matrix of $\tau$ in Fig.\,\ref{fig: tau-corr}, finding no visible departures from diagonality for both FFP10 and GRF simulations. This is particularly relevant for the FFP10 case, since these simulations contain additional foreground and non-Gaussian contributions. 

\begin{figure}
    \centering
    \includegraphics[width=0.95\textwidth]{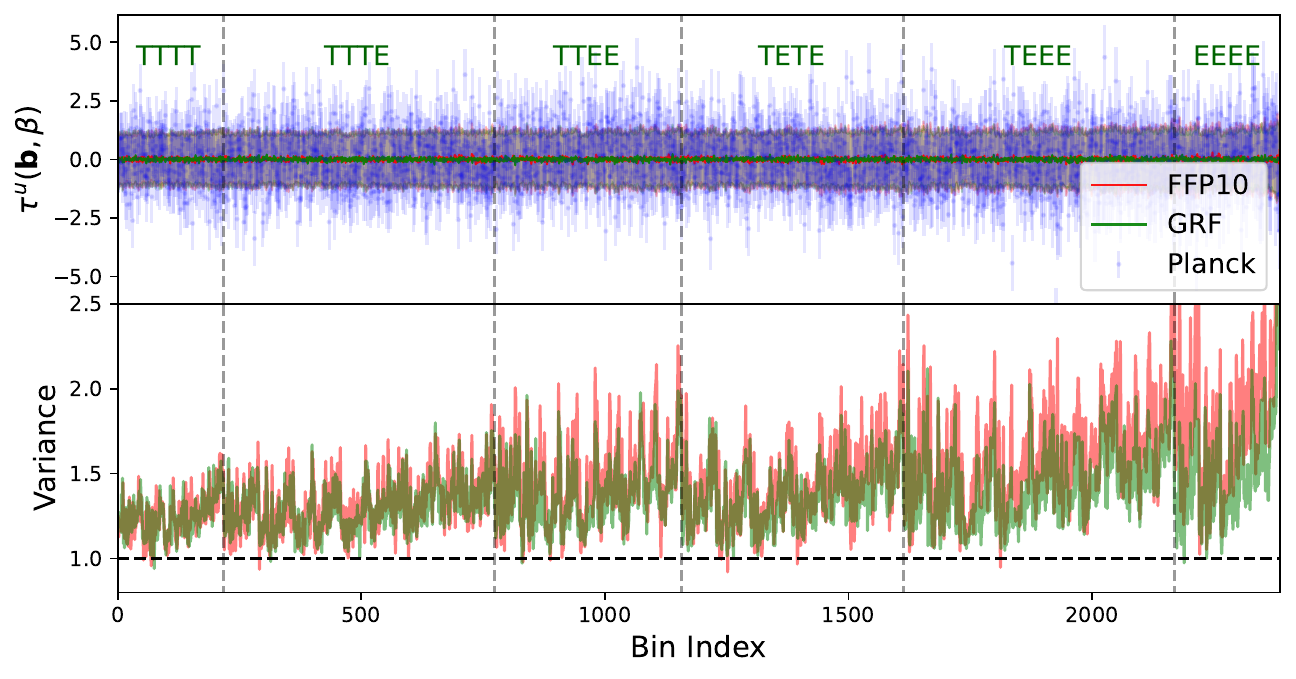}
    \caption{Rescaled parity-odd trispectrum measured from \textit{Planck} PR4 data, alongside measurements from 600 FFP10 and 1000 GRF simulations for all combinations of $T$- and $E$-modes with $4\leq \ell<518$. The top panel shows the raw measurements, whilst the bottom shows their simulated variances. The $\tau$-statistic is defined in \eqref{eq: tau-def}, with weighting chosen to decorrelate the various trispectrum bins. As in Fig.\,\ref{fig: corr}, we stack the bins into one dimension, with different polarizations indicated by dotted lines. We find no obvious detection of the trispectrum across the $2386$ bins considered herein. The variance of the simulations is larger than unity, implying that the estimators are slightly suboptimal, particularly for polarization. Whilst this does not bias our analysis, it suggests slightly stronger constraints could be wrought with a more advanced weighting scheme.}
    \label{fig: tau-data}
\end{figure}

The final ingredient in our model is the variance of $\tau$. If the trispectrum estimation pipeline is optimal and the statistic is Gaussian distributed, this should be unity (in which case $\mathcal{F}^{-1}$ and the covariance matrix match, instead of just their correlation structures). In the lower panel of Fig.\,\ref{fig: tau-data} we plot the variance of $\tau$ for the two sets of simulations, finding $\tau>1$ for most bins, with particularly pronounced signatures in bins at low $L$ and for $E$-modes. This indicates suboptimality (but not bias), and arises from (a) mask- and noise-induced correlations between pixels; (b) violations of the central limit theorem at low-$\ell$; (c) additional variance sourced by the finite number of simulations used in the estimator numerator (which are shuffled between measurements). That the signatures are different for the GRF and FFP10 simulations indicates that this is impacted also by residual non-Gaussianities induced by lensing, foregrounds, and point sources. We additionally find stronger impacts for polarized fields (with the $T$-only results matching \citep{PhilcoxCMB}, where relevant), as expected due to the more complex noise and foreground effects therein. We stress that this does not induce bias (provided the FFP10 simulations well represent \textit{Planck}); it simply implies that the eventual sensitivity of our analyses are slightly lower than those that would be obtained if the pixel covariance was perfectly well known.

Putting the above results together, we can define the following compressed statistic used to analyze the data:
\beq\label{eq: chi2-def}
    \chi^2[\hat \tau] = \sum_{u,\vec b,\beta}\frac{\left(\hat\tau^u(\vec b,\beta)-\tau^{\rm model,u}(\vec b,\beta)\right)^2}{\mathrm{var}\left[\tau^u(\vec b,\beta)\right]},
\eeq
where $\tau^{\rm model}$ is some theoretical model (possibly zero), $\hat\tau$ is the data, and $\mathrm{var}[\tau]$ is the variance extracted from simulations.\footnote{We use FFP10 simulations for this in general, though analyze the GRF distribution with the GRF variance, since these clearly differ for $E$-modes (Fig.\,\ref{fig: tau-data}).} If $\hat \tau$ is Gaussian distributed, this follows a $\chi^2$ distribution with $N_{\rm bins}$ degrees-of-freedom; if low-$\ell$ and low-$L$ modes are important in the analysis, this may be a poor assumption. As such, we will principally use the \textit{empirical} distributions of $\chi^2$ defined from simulations, thus tiptoeing into the realm of frequentist statistics and simulation based inference.

\subsection{Data}
\noindent In Fig.\,\ref{fig: tau-data} we plot the rescaled trispectrum measurements from the \textit{Planck} PR4 dataset in addition to those from the simulations. For both sets of simulations, the mean trispectrum is consistent with zero, indicating that our pipeline is not biased. The \textit{Planck} results also show no discernible signal, though the noise makes it difficult to make strong conclusions from visual inspection alone.

To robustly assess the statistical properties of our data, we perform a blind frequentist test for parity-violation. Here, we compute the $\chi^2$ statistic of \eqref{eq: chi2-def} assuming zero theory model, and compare the empirical distributions from simulations to the $\chi^2$ value obtained from \textit{Planck}. This test effectively asks the question: ``do we find \textit{any} evidence of parity-violating physics in the \textit{Planck} dataset?''. As such, it is blind to the physical shape of a parity-violating model, and one could obtain stronger constraints on particular models using a dedicated analysis (as in \S\ref{sec: results-gauge}). 

The results of this test are shown in Fig.\,\ref{fig: chi2-results}. We firstly note that the empirical PDFs are somewhat broader than the expected $\chi^2$ distribution; this suggests that the trispectrum likelihood is non-Gaussian, and implies that a classical $\chi^2$-analysis could yield biased results, whence noise fluctuations in the tails are interpreted as high-significance detections (which may partially explain the results of \citep{Hou:2022wfj}). The FFP10 and GRF distributions are generally consistent; this implies that foreground and lensing effects (present only in FFP10) are not leading to additional non-Gaussianity (though we note that the two sets of simulations have different variances, and modeling the FFP10 trispectrum with the GRF variance would lead to biased results). Turning to the \textit{Planck} data, we find a $\chi^2$ value highly consistent with both empirical distributions, with a rank of $201/600$ for FFP10 (equivalent to $-0.4\sigma$) or $340/1000$ for the GRFs.\footnote{If uses only the GRF simulations (for the empirical distribution and the $\tau$ variance entering \eqref{eq: tau-def}), the \textit{Planck} data has a rank of $906/1000$, equivalent to $1.3\sigma$.} Given that both sets of simulations are parity-conserving, this test leads to the following conclusion: \textbf{we find no evidence of scalar parity-violation in the large-scale ($4\leq \ell<518$) \textit{Planck} polarized data-set}.

Although the above analysis has returned a null result, it is important to assess the various assumptions in the measurement. To this end, we consider the following variations of the analysis in Appendix \ref{app: consistency}:
\begin{itemize}
    \item \textbf{Inclusion of $B$-modes}: This accounts for mask-induced leakage between $E$- and $B$-modes, by additionally estimating trispectra containing a single $B$-mode (e.g., $TTTB$, $TETB$), then discarding them after application of the inverse normalization matrix.
    \item \textbf{Inclusion of parity-even trispectra}: This removes any leakage between parity-odd and parity-even modes induced by the mask, analogous to the $B$-mode case.
    \item \textbf{Removal of four-field terms}: This assesses the impact of the disconnected trispectrum terms by computing only the full trispectrum, $\propto\av{\prod_{i=1}^4a_{\ell_i m_i}^{X_i}}$, which does not require any Monte Carlo estimation for the trispectrum numerator. In the ideal limit, such terms do not contribute to parity-odd spectra (though may enhance variances).
    \item \textbf{Removal of polarization}: This assesses the impact of the $E$-mode spectra, by restricting to $TTTT$-trispectra only. This additionally validates the results of \citep{PhilcoxCMB} with updated \textit{Planck} PR4 data.
\end{itemize}
The first and second procedures are somewhat more expensive than the fiducial approach, thus we restrict to 100 simulations in both analyses. In all cases, we find no evidence for parity-violation.

\section{Results: Gauge Model Test}\label{sec: results-gauge}
\noindent In \citep{PhilcoxCMB}, a variety of models were compared to the (broadly binned) \textit{Planck} parity-odd temperature trispectra to yield constraints on their amplitudes. Here, we take the most promising model (``detected'' at $\approx 2\sigma$ previously) and perform a dedicated polarized trispectrum analysis to yield tight constraints on its amplitude. We utilize the model described in \citep{Shiraishi:2016mok}, which invokes the coupling between a pseudo-scalar field $\phi$ and a gauge field $A_\mu$ to yield parity-odd trispectra via a tree-level interaction sourced by a Chern-Simons term in the inflationary Lagrangian, involving the vacuum expectation value of the gauge field. In terms of the curvature perturbation $\zeta$, the trispectrum is a special case of the general collapsed form introduced in \citep{Shiraishi:2016mok}
\beq \label{eq: general-Todd-model}
T_{\zeta}(\vk_1,\vk_2,\vk_3,\vk_4)
&=& -i\sum_{n\geq 0} d_n^{\rm odd}\left[L_n(\hk_1\cdot\hk_3)+(-1)^nL_n(\hk_1\cdot\hs)+L_n(\hk_3\cdot\hs)\right]\\\nonumber
&&\,\times\,(\hk_1\times\hk_3\cdot\hs) P_\zeta(k_1)P_\zeta(k_3)P_\zeta(s) 
+ \text{23 perms.},
\eeq 
where $\vs \equiv \vk_1+\vk_2$, $P_\zeta$ is the primordial power spectrum and $L_n$ is a Legendre polynomial of integer order. In the gauge field scenario, the amplitude of the model is set by $A_{\rm gauge} = -d_0^{\rm odd}=d_1^{\rm odd}/3$ (equal to $A_{\rm CS}$ of \citep{PhilcoxCMB}), which is proportional to the fractional energy density in the gauge field's vacuum expectation value. For generality, we will additionally consider the joint constraints on $d_{0}^{\rm odd}$ and $d_1^{\rm odd}$ in the below, though we note that the $n=1$ term is strongly dominant in the above expansion \citep[cf.][]{Shiraishi:2016mok}.

Further discussion of the model can be found in \citep{Shiraishi:2016mok}, as well as its application to large-scale structure and CMB in \citep{Cabass:2022oap,Philcox:2022hkh} and \citep{PhilcoxCMB} respectively. Here, we extend the model to the polarized sector using a sum-separable calculation described in Appendix \ref{app: model}. Whilst the mathematics is a little involved, the end-product is simple: a set of binned trispectrum amplitudes $t^{\rm th, u}(\vec b,\beta)$ that can be rescaled to $\tau$ measurements via \eqref{eq: tau-def}. Since the model is dominated by configurations with low-$L$ (analogous to the $\tau_{\rm NL}$ parity-even template), we restrict our analysis to collapsed tetrahedra, which allows for extension to higher $\ell_{\rm max}$ in reasonable computation time, as discussed in \S\ref{sec: estimation}.

\begin{figure}
    \centering
    \includegraphics[width=0.8\textwidth]{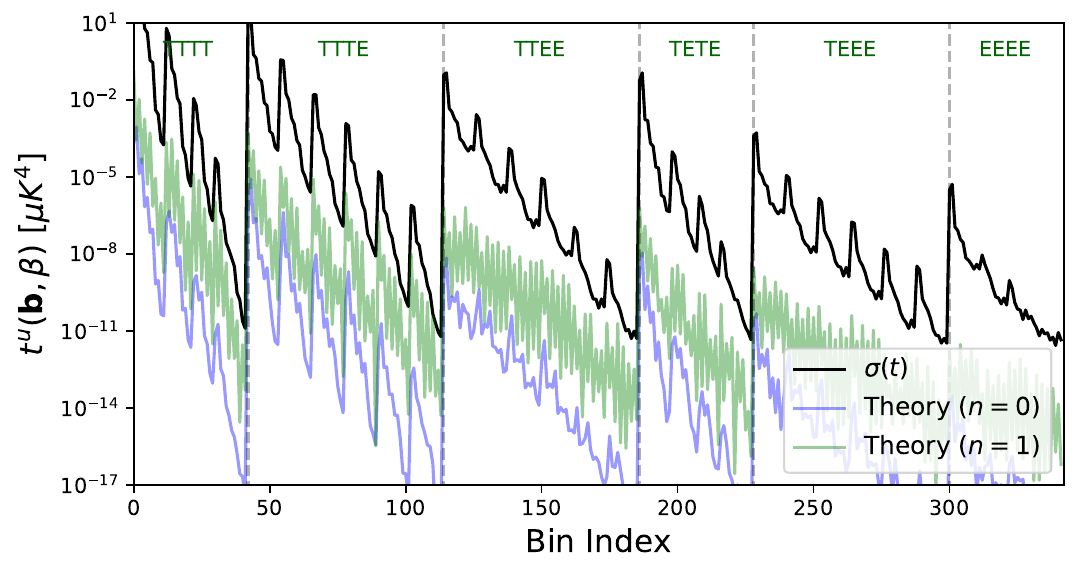}
    \caption{Theoretical models for the CMB $T$- and $E$-mode trispectra, alongside the empirical trispectrum error extracted from 100 FFP10 simulations. We show the $n=0$ and $n=1$ templates given in \eqref{eq: general-Todd-model}, the second of which well approximates the gauge field model. In contrast to previous figures, results are shown for $5\leq \ell<2000$, with internal $2\leq L<10$, where the model is large. The fiducial amplitudes of the model are set to $d_n^{\rm odd}=10^4$ for visualization (corresponding to $A_{\rm gauge}=10^4/3$), and we take the absolute value of the model correlators. Comparing theory and error, we expect that the constraints will be dominated by $TTTT$ and $TTTE$ trispectra on small scales.}
    \label{fig: tl-model}
\end{figure}

In Fig.\,\ref{fig: tl-model}, we plot the parity-odd trispectrum predicted by the first two terms in the expansion of \eqref{eq: general-Todd-model}, calculated as described in Appendix \ref{app: model}, alongside the empirical trispectrum variance from the FFP10 simulations. Being a projection of high-dimensional data, the figure is difficult to interpret: however, it is clear that the trispectrum amplitudes are largest at low-$\ell$ (particularly at low $L$), and generally suppressed in $E$-modes compared to $T$-modes. The latter occurs due to the smaller transfer function, and echoes the lower signal-to-noise found in the \textit{Planck} $EE$ power spectrum compared to $TT$. In comparison to the variance, it is clear that the strongest signatures will be found in $TTTT$ and $TTTE$ trispectra, particularly at high-$\ell$ (where the variance is suppressed). Finally, the $n=0$ and $n=1$ shapes (which differ in their angular complexity), have markedly different signatures, with the $n=1$ case (which dominates the gauge field model) having significantly larger amplitude, since it is enhanced in the collapsed limit \citep[cf.][]{Shiraishi:2016mok}.

\begin{figure}
    \centering
    \includegraphics[width=0.9\textwidth]{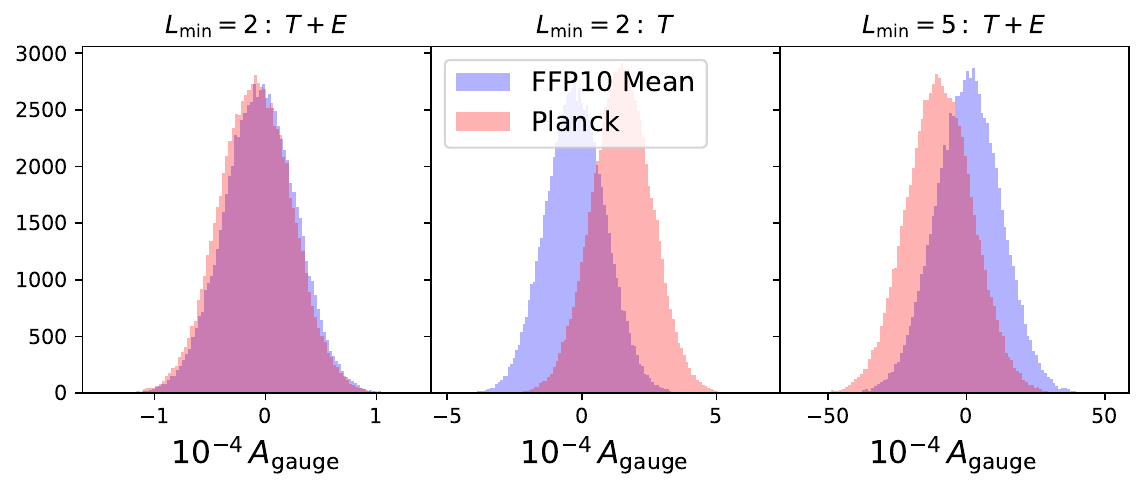}
    \caption{Constraints on the amplitude of the gauge-field model \citep{Shiraishi:2016mok} for the parity-violating trispectrum. We show the posterior for both the \textit{Planck} PR4 data (up to $\ell=2000$) and the mean of 100 FFP10 simulations. The left panel shows the fiducial analysis with internal $L\geq 2$, whilst the central and right panels respectively show the results of removing $E$-modes and restricting to $L\geq 5$. The fiducial results find $A_{\rm gauge} = (-0.9 \pm 3.3) \times 10^3$ (\textit{Planck}) and $A_{\rm gauge} = (-0.5 \pm 3.3) \times 10^3$ (FFP10), with no detection of parity-violation. Constraints degrade by a factor $\approx 3$ with the excision of $E$ modes, or by a factor $\approx 35$ with the removal of low-$L$ configurations.}
    \label{fig: gauge-pdf}
\end{figure}

\begin{figure}
\centering
\begin{minipage}{.52\textwidth}
  \includegraphics[width=0.7\textwidth]{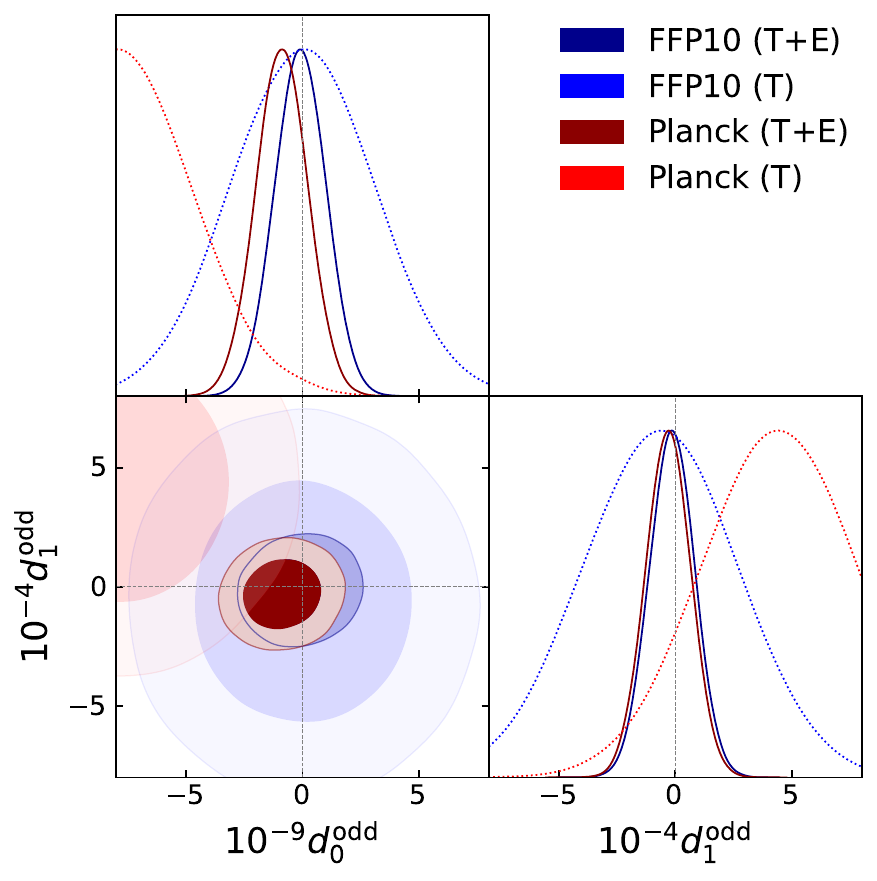}
    \caption{Constraints on the general parity-odd trispectrum given in \eqref{eq: general-Todd-model} \resub{from the mean of 100 FFP10 simulations (blue)} and \textit{Planck} data (red) up to $\ell_{\rm max}=2000$, as in Fig.\,\ref{fig: gauge-pdf}. We show \resub{two sets of contours: $T$-only (the wider and lighter contours), and $T+E$ (the darker and narrower contours)}. The best-fit \textit{Planck} constraints are $d_0^{\rm odd} = (-0.9\pm1.1)\times 10^9$, $d_1^{\rm odd}=(-3.1\pm 9.8)\times 10^3$ from temperature and polarization, or $d_0^{\rm odd}=(-7.9\pm 3.1)\times 10^9$, $d_1^{\rm odd}=(4.5\pm 3.3)\times 10^4$ from temperature alone.}
    \label{fig: joint-constraints}
\end{minipage}%
\hspace{10pt}
\begin{minipage}{.44\textwidth}
  \includegraphics[width=0.99\textwidth]{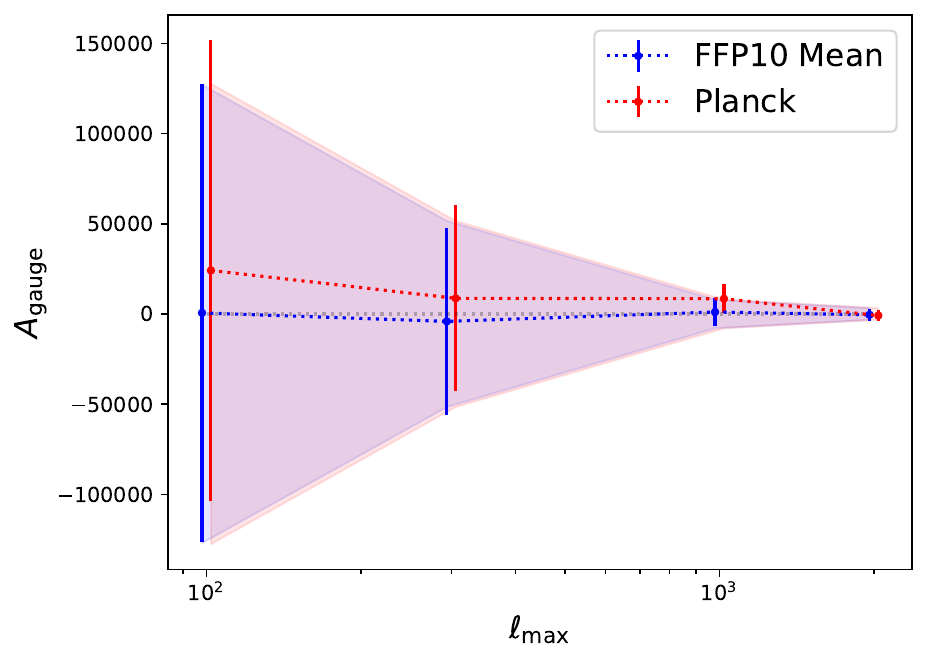}
    \caption{Dependence of the gauge-model constraints shown in Fig.\,\ref{fig: gauge-pdf} on scale cuts, via the maximum multipole $\ell_{\rm max}$. We show the $1\sigma$ error-bar both for the mean of 100 FFP10 simulations (blue) and \textit{Planck} PR4 data (red). We include both $T$- and $E$-modes in all analyses and use internal $2\leq L<10$. We find strong dependence on $\ell_{\rm max}$, though the scaling tempers slightly beyond $\ell_{\rm max}\gtrsim 1000$, likely due to projection effects.}
    \label{fig: gauge-lmax}
\end{minipage}%
\end{figure}

Using \eqref{eq: chi2-def}, we can compare the observed and theoretical trispectra, and thus place constraints on the gauge field model amplitude, $A_{\rm gauge}$ (or the model-agnostic amplitude $d_n^{\rm odd}$). For this purpose, we assume a Gaussian posterior for the amplitude parameters with an flat (pseudo-)prior of infinite extent; this is appropriate via the central limit theorem. The results of this analysis are shown in Fig.\,\ref{fig: gauge-pdf}. From the FFP10 simulation results, we verify that our pipeline is unbiased in all cases, with all the inferred value of $A_{\rm gauge}$ consistent with zero. From the \textit{Planck} data, we also find no detection of $A_{\rm gauge}$, with a $1\sigma$ contour $A_{\rm gauge}=(-0.9\pm 3.3)\times 10^3$ for the fiducial analysis. This can be translated into a constraint on the energy density of the gauge field relative to the inflaton (assuming fiducial parameters with coupling $\gamma=1$): $\rho_{\rm gauge}/\rho_\phi = (-0.7\pm 2.6)\times 10^{-19}$. As shown in Fig.\,\ref{fig: joint-constraints}, we may similarly constrain the $d_n^{\rm odd}$ parameters; the \textit{Planck} data yields $d_0^{\rm odd}=(-0.9\pm 1.1)\times 10^9$, $d_1^{\rm odd}=(-3.1\pm9.8)\times 10^3$, with a much stronger bound for the $n=1$ mode, as expected. The conclusion is clear; \textbf{we report no evidence for the parity-odd trispectrum models}.

In Figs.\,\ref{fig: gauge-pdf},\,\ref{fig: joint-constraints}\,\&\,\ref{fig: gauge-lmax}, we additionally assess the dependence of our constraints on a number of modeling choices. Firstly, we consider removing the $E$-mode data. This inflates constraints by a factor $\approx 3\times$, indicating that large-scale polarization data can be of great use when constraining primordial models. \resub{For the \textit{Planck} data, the $T$-only results show a slight ($\approx 2\sigma$) preference for $A_{\rm gauge}>0$ and $d_1^{\rm odd}>0$ (Figs.\,\ref{fig: gauge-pdf}\,\&\,\ref{fig: joint-constraints}); this is here identified as a statistical fluctuation, given that (a) the FFP10 results are unbiased and (b) the deviation vanishes when $E$-modes are included.} Secondly, we consider the removal of the first $L$-bin, giving $L\geq 5$. This degrades the constraints by a factor $\approx 35$, implying that the model is dominated by the lowest $L$-modes, as discussed in \citep{Shiraishi:2016mok}. This implies that the measurement would be highly sensitive to large-scale systematics, such as foreground-induced quadrupolar anisotropies in the power spectrum, analogously to $\tau_{\rm NL}$ constraints. Thirdly, we repeat the analysis without removing the disconnected trispectrum contributions (as discussed in \S\ref{sec: results-blind}; this is found to bias the mean-of-FFP10 results at $1\sigma$, thus we conclude that such terms are important (in the presence of inhomogeneous noise and masking). Finally, we consider the scale dependence for the external trispectrum legs in Fig.\,\ref{fig: gauge-lmax}, whence we constrain $A_{\rm gauge}$ using subsets of the full data-vector. Whilst this approach does not quite emulate a complete analysis at lower $\ell_{\rm max}$ (which would feature a finer binning), it nevertheless indicates a strong dependence of $\sigma\left(A_{\rm gauge}\right)$ on $\ell_{\rm max}$. Here, we find this to be somewhat slower than quadratic, despite the predictions of \citep{Shiraishi:2016mok,Kalaja:2020mkq} (for a $\tau_{\rm NL}$-like diagram). This is likely due to a collection of effects: (a) non-linearity in the polarization covariance on small-scales (captured in our simulated-based variances), (b) damping and projection effects at $\ell\gtrsim 1000$, (c) suboptimal binning strategies (chosen based on computational considerations). \resub{Using finer $\ell$-bins} may lead to slightly tighter bounds, though, in the absence of a definitive target for $A_{\rm gauge}$ or $d_n^{\rm odd}$, it is unclear to what extent this is useful.

\section{Discussion}\label{sec: summary}
\noindent In the above sections, we have presented novel constraints on the parity-odd sector of the Universe using \textit{Planck} temperature and polarization data, facilitated by efficient new trispectrum estimators \citep{Philcox:2023uwe,Philcox:2023psd}. It is now our duty to place these in context: what does this imply for observations and models? How can the analysis be extended? What is the outlook for the future?

The blind analysis performed in \S\ref{sec: results-blind} places competitive constraints on scalar parity-violating phenomena, both in the primordial Universe, and at late times, through CMB secondary effects. Importantly, these results do not depend on a physical model: any process with sufficiently large amplitude should show up as an excess signal in the simulation-based $\chi^2$-analysis. That said, our analysis is limited to comparatively large scales ($\ell<518$), due to the computational burden, and we cannot rule out the possibility of signals lurking at higher-$\ell$ (for example in CMB secondaries, such as lensing). 

It is interesting to consider the strength of such constraints in relation to previous CMB temperature and galaxy analyses \citep{PhilcoxCMB,Philcox:2022hkh,Hou:2022wfj}. As shown in \citep{PhilcoxCMB}, the \textit{Planck} dataset provides far stronger constraints on scale-invariant primordial models than the BOSS galaxy survey, since it contains $\approx 250\times$ more primordial modes. As noted in \citep{PhilcoxCMB}, this strongly implies that the signatures of parity-violation reported in \citep{Philcox:2022hkh,Hou:2022wfj} are not primordial in origin. The addition of $E$-modes can only boost the signal-to-noise: the precise gain, however, will depend strongly on the types of primordial model considered. At low $\ell\lesssim 5$, the number of modes in $T+E$ analyses is $\approx 80\%$ larger than those in $T$-alone (quantified as in \citep{PhilcoxCMB}, cf.\,\citep{Sailer:2021yzm} for galaxy surveys), though the ratio drops significantly at high-$\ell$, whence polarization noise dominates. As such, models with important contributions at low-$\ell$ can be expected to benefit significantly from the addition of $E$-modes. The situation, in practice, is a little more complex, since primordial physics affects $T$- and $E$-modes in different ways, with different sensitivity to noise and projection effects. As such, the true gain can be obtained only by testing individual models (or through Fisher forecasts).

In \S\ref{sec: results-gauge}, we considered one such model: the $U(1)$ gauge-field scenario of \citep{PhilcoxCMB}. In this case, inclusion of $E$-mode data significantly bolstered the constraints (by a factor $\approx 3$), with a $1\sigma$ bound of $A_{\rm gauge}=(-0.9\pm3.3)\times 10^3$, or $\rho_{\rm gauge}/\rho_\phi = (-0.7\pm 2.6)\times 10^{-19}$ for the fractional gauge field energy density. This is in accordance with the Fisher forecasts of \citep[Fig.\,4]{Shiraishi:2016mok}, but several orders of magnitude stronger than the previously reported bounds of  $A_{\rm gauge} = (-2.8\pm1.3)\times 10^7$, from \textit{Planck} temperature \citep{PhilcoxCMB}. The huge improvement over the former work is due to a number of factors: (a) extension to much smaller scales (from $\ell_{\rm max}\approx 500$ to $2000$), (b) inclusion of the lowest $L$-modes (down to $L=2$), (c) addition of $E$-mode data, (d) improved calculation of the theoretical templates, with less noise bias. The large-scale structure results from \citep{Philcox:2022hkh} appear slightly more optimistic than those found herein; as noted in \citep{PhilcoxCMB}, this is erroneous, since the former works did not account for mode decorrelation from non-linear effects, thus heavily overestimated the primordial physics content of the galaxy statistics (by around two orders of magnitude, based on mode counting). Finally, we note that, whilst we have only focused on one model in this study, the conclusions are more general: any parity-violating model of interest can now be constrained using the \textit{Planck} $T$- and $E$-mode data. Though these are already indirectly bound from the blind analyses, targeted analyses will always yield stronger constraints due to optimal weighting of the signal.

The next decade will see considerable improvements in the volume and quality of CMB data, thus there is significant hope for obtaining stronger bounds on the above models and other types of new physics. As discussed in \S\ref{sec: data}, gains on primordial physics will be driven by polarization rather than temperature, due to the cosmic variance limit, and proliferation of secondary effects in temperature such as the kinetic Sunyaev-Zel'dovich signal. As demonstrated in \citep{Kalaja:2020mkq}, there is much to gain from larger $\ell_{\rm max}$ (which is currently $\approx 600$ in polarization, due to noise limitations), though pushing beyond $\ell_{\rm max}\approx 2000$ will be hampered by projection effects. Collapsed and squeezed models show particularly significant dependence on $\ell_{\rm max}$, with predicted scalings up to $(S/N)^2\sim \ell_{\rm max}^4$ depending on the precise inflationary model considered. Even though we have detected nothing in this work, there is strong hope for the future!

\acknowledgments
{\small
\begingroup
\hypersetup{hidelinks}
\noindent We thank Will Coulton, Adri Duivenvoorden, and Masahiro Takada for insightful discussions. \resub{We are additionally grateful to the anonymous referee for insightful comments.} OHEP is a Junior Fellow of the Simons Society of Fellows. MS is supported by JSPS KAKENHI Grant Nos. JP20H05859 and JP23K03390. OHEP thanks IPMU for hosting a visit during which this work was started, as well as \href{https://www.flickr.com/photos/198816819@N07/53074044854/}{Llawrence the Llama} for spiritual guidance.

\endgroup
}

\appendix

\section{Gauge-Field Model}\label{app: model}
\noindent Here, we outline the derivation and computation of the gauge-field trispectrum model described in \citep{Shiraishi:2016mok}. This follows an analogous derivation in \citep{Philcox:2023psd}, but extended to include $E$-mode polarization, and a more efficient factorization scheme. We begin with the relation the CMB harmonic coefficients and the primordial curvature trispectrum, $T_\zeta$:
\beq\label{eq: Tl-T-curvature}
    \bigg\langle\prod_{i=1}^4a^{X_i}_{\ell_im_i}\bigg\rangle_c &=& (4\pi)^4\left[\prod_{i=1}^4i^{\ell_i}\int_{\vk_i}\mathcal{T}^{X_i}_{\ell_i}(k_i)Y^*_{\ell_im_i}(\hk_i)\right]T_\zeta(\vk_1,\vk_2,\vk_3,\vk_4)\delD{\vk_{1234}},
\eeq
for curvature transfer function $\mathcal{T}_\ell^X$, with only $\mathcal{T}_\ell^B=0$ at leading order. Following \citep{Philcox:2023psd}, we can write the unbinned trispectrum (the RHS of \eqref{eq: Tl-T-curvature}) as
\beq
    T^{\ell_1\ell_2\ell_3\ell_4,u}_{m_1m_2m_3m_4} &=& -\sqrt{2}i^{\ell_{1234}}A_{\rm gauge}(\Delta_\zeta^2)^3(4\pi)^{11/2}
    \sum_{L_1L_2L_3L_4L'L''}i^{L_{1234}-L'+L''}\mathcal{C}_{L_1L_2L_3L_4L'L''}\tjo{L_1}{L_2}{L'}\tjo{L_3}{L_4}{L''}\nonumber\\
    &&\qquad\times\,\int x^2dx\,\int y^2dy\,h^{\ell_1L_1,u_1}_{-3}(x)h^{\ell_2L_2,u_2}_{0}(x)h^{\ell_3L_3,u_3}_{-3}(y)h^{\ell_4L_4,u_4}_{0}(y)f_{L'L''}(x,y)\nonumber\\
    &&\qquad\times\,\int d\hs\,\prod_{i=1}^4\left[\int d\hk_iY^*_{\ell_im_i}(\hk_i)\right]\P_{L_1L_2L'}(\hk_1,\hk_2,\hs)\P_{L_3L_4L''}(\hk_3,\hk_4,\hs)\nonumber\\
    &&\qquad\times\,\sum_{\lambda_1\lambda_3\lambda}A_{\lambda_1\lambda_3\lambda}\P_{\lambda_1\lambda_3\lambda}(\hk_1,\hk_3,\hs)+\text{23 perms.},
\eeq
where $u$ is a quadruplet of fields. This uses the definitions 
\beq
    h^{\ell L,X}_{\alpha}(x)&=&\int\frac{k^2dk}{2\pi^2}k^\alpha\mathcal{T}^X_{\ell}(k)j_L(kx), \qquad f_{LL'}(x,y) =\int\frac{s^2ds}{2\pi^2}\frac{j_L(sx)j_{L'}(sy)}{s^3}
\eeq
with $A_{111}=1$, $A_{221}=-A_{212}=A_{122}=1/\sqrt{5}$ and zero else.\footnote{The general trispectrum model of \eqref{eq: general-Todd-model} can be computed via the replacements $A_{\rm gauge}A_{111}\to -d_0^{\rm odd}$, and $A_{\rm gauge}A_{221}=-A_{\rm gauge}A_{121}=A_{\rm gauge}A_{122}=d_1^{\rm odd}/(3\sqrt{5})$.} Furthermore, $\mathcal{C}_{\ell_1\cdots\ell_n} = \sqrt{(2\ell_1+1)\cdots(2\ell_n+1)}$ and $\P$ are the isotropic basis functions described in \citep{2020arXiv201014418C}, trivially related to the tripolar spherical harmonics \citep{1988qtam.book.....V}. Performing the angular integral as for the scalar case, we find
\beq\label{eq: tl-CS-unbinned}
    t^{u}_{\ell_1\ell_2,\ell_3\ell_4}(L)\tj{\ell_1}{\ell_2}{L}{-1}{-1}{2}\tj{\ell_3}{\ell_4}{L}{-1}{-1}{2} &=& -\sqrt{2}A_{\rm gauge}(\Delta_\zeta^2)^3(4\pi)^{5}
    \sum_{L_1L_3L'L''}i^{-\ell_1+\ell_3-L_1+L_3+L'+L''}\mathcal{C}^2_{L_1L_3L'L''}\\\nonumber
    &&\,\times\,\sum_{\lambda_1\lambda_3\lambda}A_{\lambda_1\lambda_3\lambda}\mathcal{C}_{\lambda_1\lambda_3\lambda}\begin{Bmatrix} L&\ell_1&\ell_2\\L_1&L'&\lambda_1\end{Bmatrix}\begin{Bmatrix}\lambda&\lambda_1&\lambda_3\\L&L''&L'\end{Bmatrix}\begin{Bmatrix}\ell_3&\ell_4&L\\L''&\lambda_3&L_3\end{Bmatrix}\\\nonumber
    &&\,\times\,\int x^2dx\,\int y^2dy\,h^{\ell_1L_1,u_1}_{-3}(x)h^{\ell_2\ell_2,u_2}_{0}(x)h^{\ell_3L_3,u_3}_{-3}(y)h^{\ell_4\ell_4,u_4}_{0}(y)f_{L'L''}(x,y)\\\nonumber
    &&\,\times\,\tjo{L_1}{\ell_2}{L'}\tjo{L_3}{\ell_4}{L''}\tjo{L_1}{\lambda_1}{\ell_1}\tjo{L'}{\lambda}{L''}\tjo{\lambda_3}{L_3}{\ell_3},
\eeq
which can be partially factorized into $(\ell_1,\ell_2,u_1,u_2)$ and $(\ell_3,\ell_4,u_3,u_4)$ contributions, linked by integrals and $6j$ symbols.

To compare theory and data, we must appropriately bin the theory model. Following \citep{Philcox:2023psd}, the binned \resub{trispectra} are defined as:
\beq
    \widehat{t}^{u}(\vec b,\beta) &\propto& \frac{8}{\Delta_4^u(\vec b)}\sum_{\ell_iL}\Theta_{\ell_1}(b_1)\cdots\Theta_{\ell_4}(b_4)\Theta_L(\beta)(2\ell_1+1)\cdots(2\ell_4+1)(2L+1)\\\nonumber
     &&\,\times\,\tj{\ell_1}{\ell_2}{L}{-1}{-1}{2}^2\tj{\ell_3}{\ell_4}{L}{-1}{-1}{2}^2\sum_{u'}S^{-1,u_1u_1'}_{\ell_1}\cdots S^{-1,u_4u_4'}_{\ell_4}t_{\ell_1\ell_2,\ell_3\ell_4}^{u'}(L),
\eeq
dropping constant factors, and asserting that the spectra is parity-odd. Here, the $u'$ sum is over all $2^4$ quadruplets of fields (not just the ordered pairs appearing in the trispectra of interest), and the forms are normalized by the diagonal-in-$\beta$ matrix
\beq
    \mathcal{N}^{uu'}(\vec b,\vec b',\beta) &\equiv&  
     \frac{1}{\Delta_4^u(\vec b)\Delta_4^{u'}(\vec b')}\sum_{\ell_iL}\Theta_{\ell_1}(b_1)\cdots\Theta_{\ell_4}(b_4)\Theta_L(\beta)(2\ell_1+1)\cdots(2\ell_4+1)(2L+1)\\\nonumber
     &&\,\times\,\tj{\ell_1}{\ell_2}{L}{-1}{-1}{2}^2\tj{\ell_3}{\ell_4}{L}{-1}{-1}{2}^2\left[S^{-1,u_1u_1'}_{\ell_1}\cdots S^{-1,u_4u_4'}_{\ell_4}\delta^{\rm K}_{b_1b_1'}\cdots \delta^{\rm K}_{b_4b_4'}+\text{7 perms.}\right].
\eeq
This correlates only pairs of bins where $\vec b'$ is a permutation of $\vec b$. The normalization is high-dimensional but can be efficiently computed by factorizing:
\beq
    \mathcal{N}^{uu'}(\vec b,\vec b',\beta) &\equiv&  
     \frac{1}{\Delta_4^u(\vec b)\Delta_4^{u'}(\vec b')}\sum_p\sum_L\Theta_L(\beta)(2L+1)\\\nonumber
     &&\qquad\,\times\,\left(\sum_{\ell_1\ell_2}\Theta_{\ell_1}(b_1)\Theta_{\ell_2}(b_2)(2\ell_1+1)(2\ell_2+1)\tj{\ell_1}{\ell_2}{L}{-1}{-1}{2}^2S^{-1,u_1u'_{p_1}}_{\ell_1}S^{-1,u_2u'_{p_2}}_{\ell_2}\delta^{\rm K}_{b_1b'_{p_1}}\delta^{\rm K}_{b_2b'_{p_2}}\right)\\\nonumber
     &&\qquad\,\times\,\left(\sum_{\ell_3\ell_4}\Theta_{\ell_3}(b_3)\Theta_{\ell_4}(b_4)(2\ell_3+1)(2\ell_4+1)\tj{\ell_3}{\ell_4}{L}{-1}{-1}{2}^2S^{-1,u_3u'_{p_3}}_{\ell_3}S^{-1,u_4u'_{p_4}}_{\ell_4}\delta^{\rm K}_{b_3b'_{p_3}}\delta^{\rm K}_{b_4b'_{p_4}}\right),
\eeq
where $p$ are the eight permutations above of $\{1,2,3,4\}$.\footnote{This approach was not considered in \citep{Philcox:2023psd}, leading to the prohibitive $\mathcal{O}(\ell_{\rm max}^5)$ scaling claimed previously.} Inserting \eqref{eq: tl-CS-unbinned}, the binned trispectrum numerator can be written in factorized form: 
\beq
    t^{u}(\vec b,\beta)
    &\propto&\,-\sqrt{2}A_{\rm gauge}(\Delta_\zeta^2)^3(4\pi)^5\frac{8}{\Delta_4^u(\vec b)}\sum_{\lambda_1\lambda_3\lambda}(-1)^{\lambda_3}A_{\lambda_1\lambda_3\lambda}\mathcal{C}_{\lambda_1\lambda_3\lambda}\\\nonumber
    &&\,\times\,\sum_{LL'L''}\Theta_L(\beta)(2L+1)(2L'+1)(2L''+1)\tjo{L'}{\lambda}{L''}\begin{Bmatrix}\lambda&\lambda_1&\lambda_3\\L&L''&L'\end{Bmatrix}\\\nonumber
    &&\,\times\,\frac{1}{2}\int\frac{s^2ds}{2\pi^2}\frac{1}{s^3}\left[f_{LL'\lambda_1}^{u_1u_2}(b_1,b_2;s)f_{LL''\lambda_3}^{u_3u_4}(b_3,b_4;s)-\overline{f}_{LL'\lambda_1}^{u_1u_2}(b_1,b_2;s)\overline{f}_{LL''\lambda_3}^{u_3u_4}(b_3,b_4;s)\right]
  ,
\eeq
defining
\beq
    f^{u_1u_2}_{LL'\lambda_1}(b_1,b_2;s) &=& \sum_{\ell_1\ell_2L_1}\Theta_{\ell_1}(b_1)\Theta_{\ell_2}(b_2)(2\ell_1+1)(2\ell_2+1)i^{-\ell_1-L_1+L'}(2L_1+1)\\\nonumber
    &&\,\times\,\tjo{L_1}{\ell_2}{L'}\tjo{L_1}{\lambda_1}{\ell_1}\tj{\ell_1}{\ell_2}{L}{-1}{-1}{2}\begin{Bmatrix} L&\ell_1&\ell_2\\L_1&L'&\lambda_1\end{Bmatrix}\\\nonumber
    &&\,\times\,\int x^2dx\,\tilde{h}^{\ell_1L_1,u_1}_{-3}(x)\tilde{h}^{\ell_2\ell_2,u_2}_{0}(x)j_{L'}(sx)
\eeq
where $\tilde{h}^{\ell L,u}_\alpha(y) \equiv \sum_{u'}S^{-1}_{uu'}h^{\ell L,u'}_\alpha(y)$. $\overline{f}$ is analogous but with an extra factor of $(-1)^{\ell_1+\ell_2}$ or $(-1)^{\ell_3+\ell_4}$ (to impose an odd $\ell_{1234}$ sum). 
In practice, the various integrals (appearing in $h$, $f$ and $t$) integrals are evaluated using numerical quadrature, with linearly spaced arrays in $\{k_i/\ell_i,x,s\}$, with a total of $\{5000, 2000, 1000\}$ points in each (noting that the $h$ integrals can all be precomputed, thus a dense $k$-grid is possible). Various modifications to the finite integration grids were considered to ensure that any systematic errors induced are small. 

\section{Consistency Checks}\label{app: consistency}
\noindent In this appendix, we present various consistency checks of the blind analysis. In each case, we will reproduce Fig.\,\ref{fig: chi2-results} with the modified assumptions as described in the last paragraph of \S\ref{sec: results-blind}. For consistency, we compare the $B$-mode and even-parity runs against fiducial analyses using $100$ GRF and FFP10 simulations only. We note that the associated increased noise in the variance of $\tau$ leads to a slight upwards shift in the empirical $\chi^2$ posteriors in all cases, with a fiducial \textit{Planck} parity-violation probability of $0.06\sigma$.

\begin{figure}
    \subfloat[Adding $B$-modes]{\includegraphics[width=0.48\textwidth]{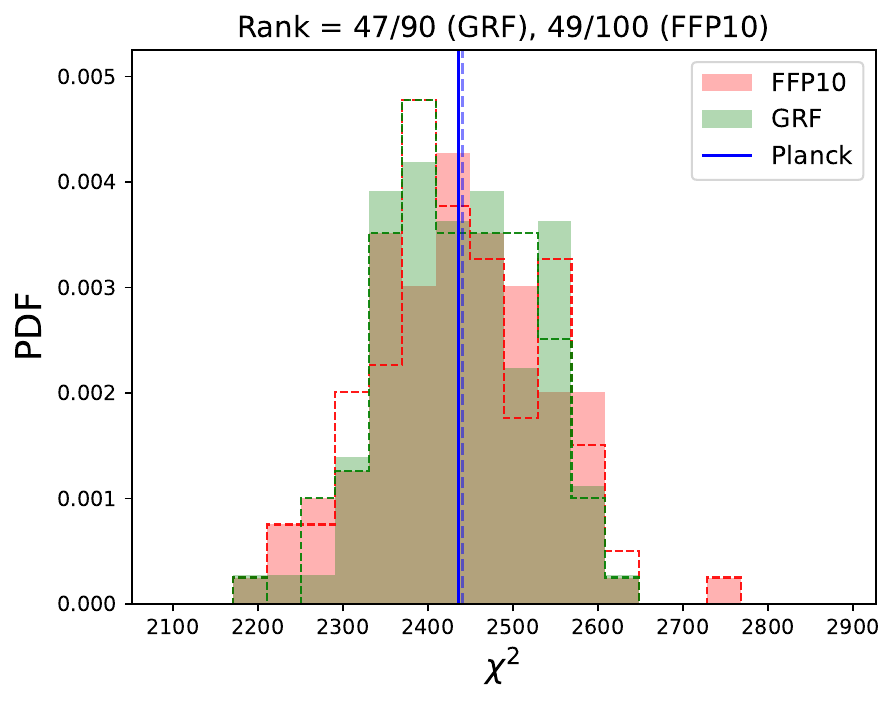}}
    \subfloat[Adding parity-even modes]{\includegraphics[width=0.48\textwidth]{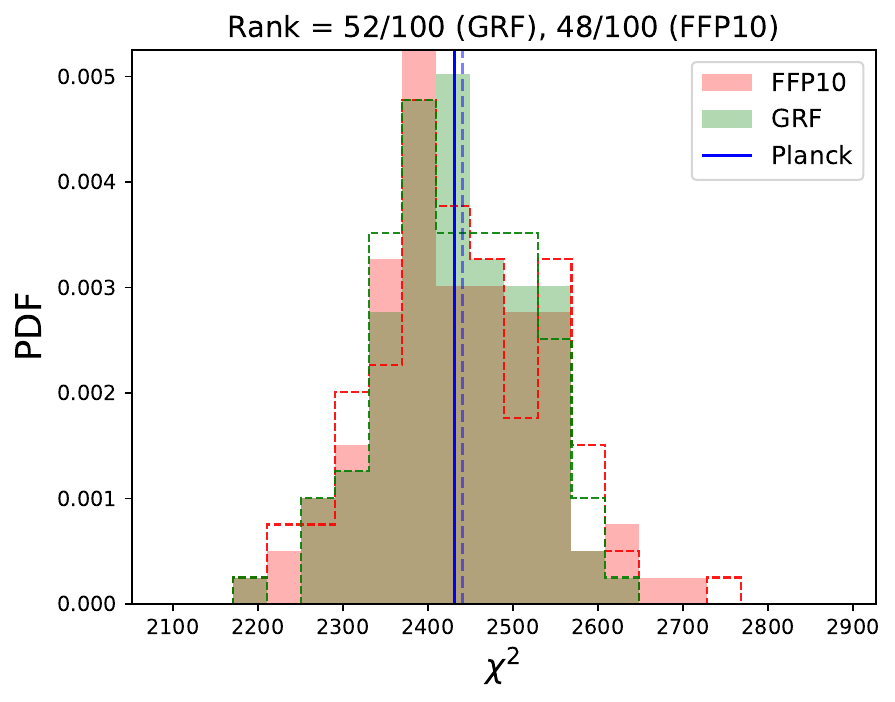}}\\
    \subfloat[Without disconnected subtraction]{\includegraphics[width=0.48\textwidth]{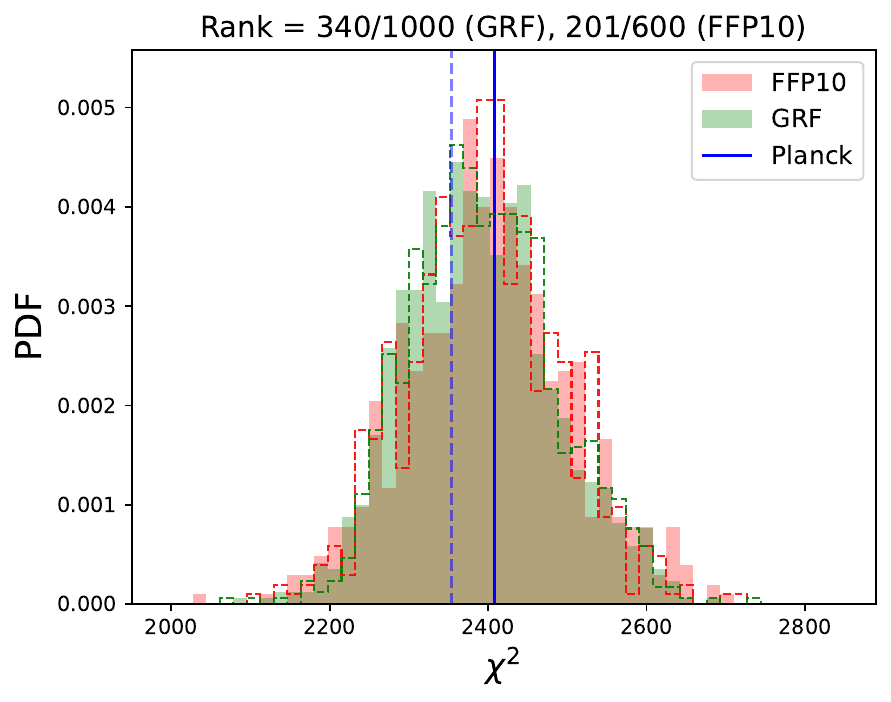}}
    \subfloat[Without $E$-modes]{\includegraphics[width=0.48\textwidth]{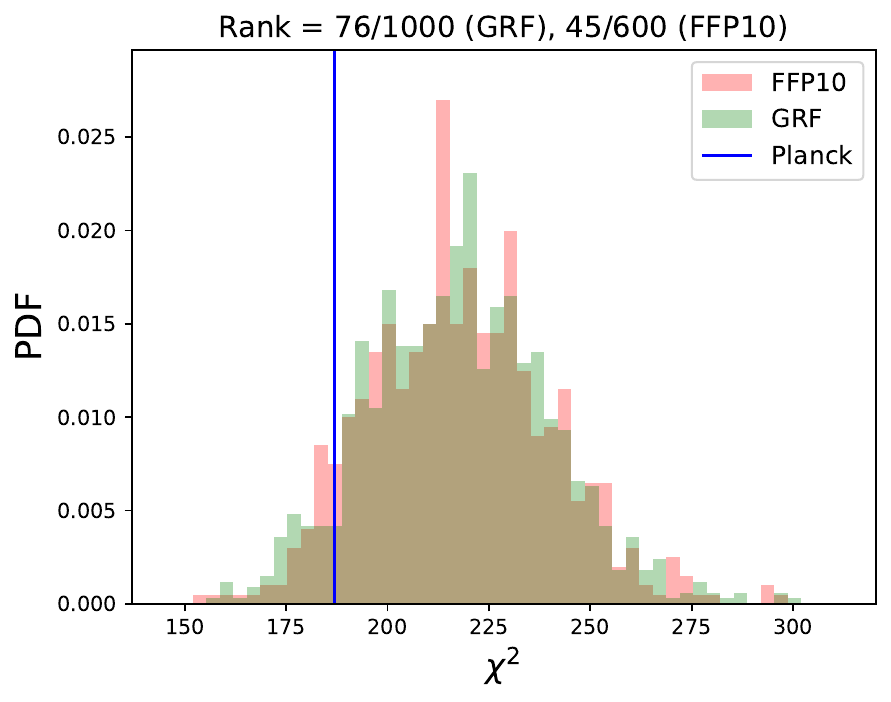}}
    \caption{As Fig.\,\ref{fig: chi2-results}, but assessing the consistency of our analyses with various modifications to our fiducial assumptions. In the first three cases (described in Appendix \ref{app: consistency}), we show the $\chi^2$ distributions from the analysis alongside the fiducial results in dashed lines; in the fourth case, we show only the new results, since the dimensionality differs. The first two tests use $100$ FFP10 and GRF simulations, whilst the second two use the full fiducial sample. In all cases, the blind tests show no evidence for parity-violation.}\label{fig: consistency}
\end{figure}

\vskip 4pt
\paragraph*{Adding $B$-modes} As described in \S\ref{sec: results-blind}, we test for leakage between $E$- and $B$-modes (and lensing effects) by additionally including $TTTB$, $TTEB$, $TETB$, $TEEB$, and $EEEB$ trispectra in our measurement pipeline (leading to a total of $8\,360$ bins), then discarding such configurations after normalization by $\mathcal{F}^{-1}$. Assuming the leakage to be small, such spectra (with one $E\to B$ replacement) dominate the signal. This roughly doubles the computation cost of trispectra, thus is avoided in the main analysis. From Fig.\,\ref{fig: consistency}a, we find no significant changes to our main results, except for slight fluctuations in the $\chi^2$ of any given simulation, due to the associated modifications to the trispectrum noise. We find no evidence for parity-violation, with the \textit{Planck} data consistent with FFP10 at $-0.01\sigma$ (from the rank tests).

\vskip 4pt
\paragraph*{Adding parity-even modes} Leakage between parity-odd and parity-even modes (odd and even $\ell_1+\ell_2+\ell_3+\ell_4$) is assessed in a similar manner by jointly estimating all trispectrum configurations (a total of $6\,260$), then removing the latter after the (decorrelating) normalization matrix is applied (which is block diagonal in the ideal case). From Fig.\,\ref{fig: consistency}b, we again find no significant change to the posterior, with smaller noise fluctuations in this case. The \textit{Planck} data has a overall probability of $-0.03\sigma$, and we find no evidence for parity-violation. 

\vskip 4pt
\paragraph*{Removing disconnected corrections} In Fig.\,\ref{fig: consistency}c we assess the impact of the disconnected terms in the trispectrum numerator. In the ideal case, these will not modify the signal, but may change the covariance of certain bins (specifically, those with two or four equal $\ell$-bins). Since this study does not require new analysis, we can use all the previously computed fiducial simulations; from the figure, we find that removal of such terms has very little effect, implying that leakage into the disconnected contributions is small in this case (though more relevant for the gauge-field results of \S\ref{sec: results-gauge}). The corresponding detection significance (relative to FFP10) remains at the fiducial value of $-0.4\sigma$.

\vskip 4pt
\paragraph*{Removing polarization}
Finally, we consider the effects of removing $E$-modes from our analysis. This emulates the results of \citep{PhilcoxCMB}, except with slightly broader binning, updated \textit{Planck} PR4 measurements, and including the $T$-$E$ correlations in the measurement pipeline (by restricting to $T$ modes only after the trispectrum is normalized). In this case, the PDF (shown in Fig.\,\ref{fig: consistency}d), is shifted to much lower $\chi^2$ values, since we reduce the dimensionality from $2386$ to $217$ bins. Once again, we find no evidence for parity-violation, with the \textit{Planck} data consistent with the FFP10 simulations at $-1.43\sigma$ in a rank test. This is expected, and shows that our new pipeline yields analogous results to the former.

\bibliographystyle{apsrev4-1}
\bibliography{refs}

\end{document}